\documentclass[%
 reprint,
superscriptaddress,
 amsmath,amssymb,
 aps,
pra,
]{revtex4-2}

\usepackage{capt-of}

\usepackage{graphicx}
\usepackage{dcolumn}
\usepackage{bm}
\usepackage{hyperref}

\begin{document}

\preprint{APS/123-QED}

\title[LAOStrain response of carbon black-polymer hydrogels]{LAOStrain response of carbon black-polymer hydrogels:\\
insights from rheo-TRUSAXS and rheo-electric experiments}
\author{Gauthier Legrand}
\affiliation{ENSL, CNRS, Laboratoire de physique, F-69342 Lyon, France}
\author{Guilhem P. Baeza}%
\affiliation{Universit\'e Claude Bernard Lyon 1, INSA Lyon, Universit\'e Jean Monnet, CNRS, UMR 5223, Ing\'enierie des Mat\'eriaux Polym\`eres, F-42023 Saint-Etienne Cedex, France}
\author{William Ch\`evremont}
\affiliation{ESRF, The European Synchrotron, 71 Avenue des Martyrs, CS40220, 38043 Grenoble Cedex 9, France\looseness=-1}
 \author{S\'ebastien Manneville}
\affiliation{ENSL, CNRS, Laboratoire de physique, F-69342 Lyon, France}
\affiliation{Institut Universitaire de France (IUF)}
 \author{Thibaut Divoux}
\affiliation{ENSL, CNRS, Laboratoire de physique, F-69342 Lyon, France}
\affiliation{International Research Laboratory, French American Center for Theoretical Science, CNRS, KITP, Santa Barbara, USA\looseness=-1}

\date{\today}

\begin{abstract}
Colloid-polymer hydrogels are commonly encountered in various applications, from flow batteries to drug delivery. Here, we investigate hydrogels composed of hydrophobic colloidal soot particles --carbon black (CB)-- and carboxymethylcellulose (CMC), a food-grade polymer functionalized with hydrophobic groups that bind physically to CB particles. As previously described in [Legrand \textit{et al.}, \textit{Macromolecules} \textbf{56}, 2298-2308 (2023)], CB-CMC hydrogels exist in two flavors:  either electrically conductive when featuring a percolated network of CB particles decorated by CMC, or insulating where isolated CB particles act as physical cross-linkers within the CMC matrix. 
Here, we compare the yielding transition of these two types of CB-CMC hydrogels under Large Amplitude Oscillatory Shear (LAOS), combining rheometry with Time-Resolved Ultra-Small-Angle X-ray Scattering (TRUSAXS) and electrical conductivity measurements. Both types of hydrogels exhibit a ``type III'' yielding scenario, characterized by an overshoot in $G''$ and a monotonic decrease in $G'$, although the underlying microscopic mechanisms differ markedly. Conductive CB-CMC hydrogels display a yield strain $\gamma_y\simeq 6\%$ concomitant with a drop in DC conductivity, indicative of the macroscopic rupture of the percolated CB network at length scales larger than a few microns, beyond the USAXS resolution. At larger strain amplitudes, the conductivity of the fluidized sample increases again, exceeding its initial value, consistent with the shear-induced formation of a transient, dynamically percolated network of CB clusters.
In contrast, insulating CB-CMC hydrogels exhibit a significantly larger yield strain, $\gamma_y \simeq 60\%$, beyond which the sample flows and the average distance between CB particles decreases, as revealed by rheo-TRUSAXS. This structural reorganization is concomitant with a more than tenfold increase in conductivity, although it remains below that of the conductive hydrogels at rest. 
\end{abstract}

\maketitle

\section{\label{sec:intro} Introduction}

In the realm of soft materials, colloidal-polymer dispersions serve as fundamental model systems for understanding self-assembly and phase behavior while also being prevalent in a wide range of applications, from drug delivery to flow batteries and direct-write assembly \cite{Schexnailder:2009,Appel:2015,Richards:2016,Costa:2020,Meslam:2022,Smay:2002,Richards:2024}. Their microstructure and mechanical properties are primarily governed by the nature of colloid-polymer interactions.  

Indeed, when polymers are non-absorbing, the colloids experience entropic depletion interactions, which can drive phase separation, gelation, or crystallization \cite{Poon:1995,Bergenholtz:2003,Laurati:2009}. The resulting gels have been extensively used as model systems to investigate the shear-induced solid-to-liquid transition commonly observed in physical gels \cite{Koumakis:2011,Sprakel:2011,Laurati:2011,Chan:2012,Moghimi:2021}, and to probe how shear history --and more broadly, the gelation pathway-- affects the structural and mechanical properties of gels at rest \cite{Koumakis:2015,Moghimi:2017,Colombo:2025}. 

When polymers can adsorb onto colloids, the phase diagram is even richer. Polymers can completely cover the surface of the particles, resulting in stable colloidal suspensions \cite{Dolan:1974,Garcia:2020}. By contrast, in the case of partial coverage, adsorbing polymers can form bridges between colloids \cite{Iler:1971}, leading to various flocculation scenarios and a wealth of time-dependent microstructures, including space-spanning networks \cite{Otsubo:1990,Gallegos:2023}. The latter gels form spontaneously at relatively low polymer concentrations, and their properties can be tuned by varying the particle size and the polymer molecular weight. In the case of long polymer chains, each macromolecule can adsorb on multiple particles, binding them into clusters whose shape depends on the electrostatic charge of the particle \cite{Wong:1992,Spalla:1993}. Under shear or vigorous shaking, these mixtures display a noticeable shear-thickening transition that results from the shear-induced aggregation of the clusters. These ``\textit{shake gels}'' display a broad range of lifespan ranging from short-lived \cite{Zebrowski:2003,Pozzo:2004,Kamibayashi:2008,Banerjee:2024b} to permanent \cite{Otsubo:1984,Otsubo:1990b,Cabane:1997}. Finally, in the limit of high polymer concentrations, colloids act as fillers within a continuous polymer matrix. In this regime, the filler-matrix interactions can be tuned to enhance the mechanical properties of the hydrogel composite \cite{ Dehne:2017,Dellatolas:2023,Gulluche:2023}.

This brief overview of the literature highlights the versatility of combining colloids and polymers to design colloidal gels via polymer bridging. Although the relationship between gel composition and linear viscoelastic properties has been systematically explored in only a few cases, key examples highlight the potential of this approach. For instance, silica colloids mixed with oppositely charged polyelectrolytes, such as poly(ethylene imine), form gels whose elasticity varies non-monotonically with the polymer content at a constant colloid volume fraction \cite{Pickrahn:2010}. Conversely, gels made of polyacrylamide and various amounts of silica nanoparticles exhibit a continuous evolution in their linear viscoelastic spectrum. This trend was leveraged to build master curves \cite{Adibnia:2017}, unraveling time-concentration superposition principles that are consistently observed with other types of polymers \cite{Pashkovski:2003}.

In a recent work \cite{Legrand:2023}, we have investigated hydrogels made of carbon black (CB) hydrophobic particles and carboxymethylcellulose (CMC), a polyelectrolyte derived from cellulose and also known as cellulose gum \cite{Keller:2020}. The backbone of CMC consists of anhydroglucose units, which contain up to three hydroxyl groups substituted by a carboxymethyl group. For low degrees of substitution $\rm DS$, typically $\rm DS \lesssim 0.9$ \cite{Lopez:2018}, the substitution occurs heterogeneously along the polymer chains, which display unsubstituted hydrophobic regions \cite{Debutts:1957}. Due to their hydrophobic nature, the CB particles naturally bind to the CMC hydrophobic patches, forming hydrogels over a broad range of CB content and CMC concentrations \cite{Legrand:2023,Park:2023,Herveou:2025}.

Conducting extensive rheological measurements and electrical impedance spectroscopy, we have shown that CB-CMC hydrogels come in two flavors: electrically conductive or insulating, which correspond to two different types of microstructures and viscoelastic properties \cite{Legrand:2023}. On the one hand, conductive CB-CMC hydrogels comprise a percolated network of CB particles, which is stabilized in water by the CMC polymer. These conductive hydrogels display a glassy-like viscoelastic spectrum that is merely frequency-dependent. On the other hand, insulating CB-CMC hydrogels consist of a polymer matrix in which the CB particles act as physical crosslinkers. As a result, the viscoelastic spectrum is frequency-dependent, in a way that strongly varies with the particle and polymer content. Yet, across a broad range of compositions, all spectra can be rescaled onto a master curve following a time-concentration superposition principle. 
Finally, the transition between these two types of hydrogels is controlled by a critical mass ratio $r=m_{\rm CMC}/m_{\rm CB}=r_c$ below, which CB-CMC hydrogels are conductive and above which they are insulating. We found that $r_c \simeq 0.037$, which corresponds to about $5$ CMC polymer per CB particle \cite{legrand:tel-04768650}.

In the present article, we report on the non-linear rheological response of these two types of CB-CMC hydrogels. We focus on their shear-induced yielding process under strain-controlled oscillatory shear of increasing amplitude, hereafter referred to as Large Amplitude Oscillatory Shear Strain (LAOStrain) tests. To gain additional insights into the microscopic scenario underpinning the yielding transition, two complementary techniques were coupled with rheometry, namely time-resolved electric measurements and Time-Resolved Ultra-Small-Angle X-ray Scattering (TRUSAXS) experiments.  

We show that conductive and insulating CB-CMC hydrogels show qualitatively similar responses in their viscoelastic response to LAOStrain tests, characterized by a monotonic decrease in $G'$, an overshoot in $G''$, and a crossover of $G'$ and $G''$ associated with the sample yielding. Yet, these responses display striking different quantitative features that are well separated by the critical mass ratio $r_c$ previously identified by comparing their linear viscoelastic properties and electrical properties at rest \cite{Legrand:2023}. Additionally, time-resolved rheo-electric tests show that conductive CB-CMC hydrogels display a drop in conductivity beyond the yield strain, which points to a shear-induced failure of the CB network involving a brittle-like, localized failure scenario. Indeed, the disruption of the CB network is beyond the TRUSAXS resolution, which shows that the CB network must fragment into large-scale clusters. Further increasing the strain oscillation amplitude eventually leads to the formation of a transient, dynamic percolated network of CB clusters, which disappears rapidly upon flow cessation. 
In contrast, rheo-TRUSAXS measurements on insulating CB-CMC hydrogels reveal significant rearrangements at the particle scale, with a decrease in the interparticle distance. Time-resolved rheo-electric experiments further show that beyond the overshoot in $G''$, the sample conductivity displays a tenfold increase, albeit without reaching the conductivity of conductive CB-CMC hydrogels.

These findings constitute the core of the present article, which is organized as follows. Section~\ref{sec:materials} describes the experimental techniques used to study CB-CMC hydrogels. In Section~\ref{sec:Results0}, we report on Ultra-Small-Angle X-ray Scattering (USAXS) measurements in CB-CMC dispersions with varying compositions, confirming the existence of two distinct types of CB-CMC hydrogels, previously identified through linear viscoelastic measurements and Electrical Impedance Spectroscopy \cite{Legrand:2023}. Section~\ref{sec:Results1} focuses on the non-linear rheological behavior of these hydrogels during LAOStrain tests. The results of these tests are interpreted by intra-cycle analysis and further supported by time-resolved rheo-electric and rheo-TRUSAXS experiments. Finally, Section~\ref{sec:Discussion} offers a short discussion and concluding remarks.

\section{\label{sec:materials} Materials and Methods}
\subsection{\label{subsec:sample_prep} Samples preparation}
\begin{figure}[!th]
    \centering
\includegraphics[width=0.9\linewidth]{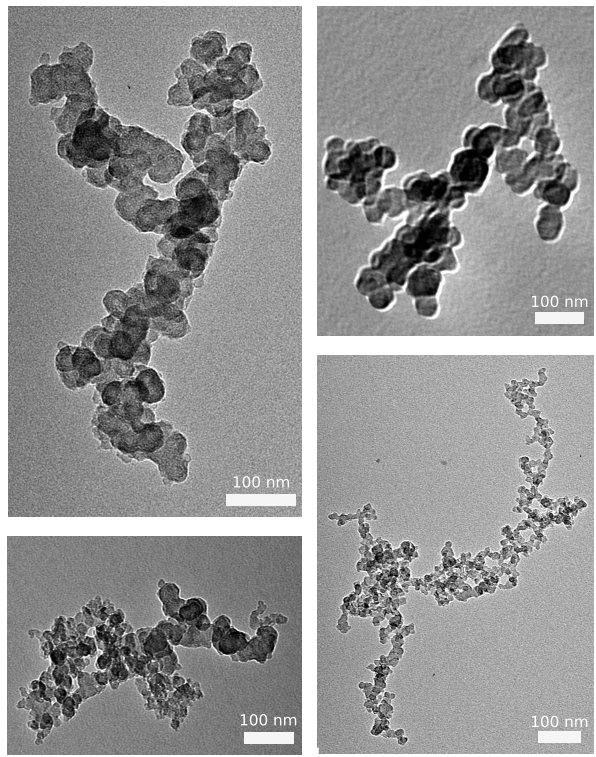}
    \caption{Representative Transmission Electron Microscopy (TEM) images of individual carbon black (Vulcan XC72R) particles. The white scale bar in each image corresponds to $100~\rm nm$. Samples were prepared by depositing a $5~\rm \mu L$ drop of a dilute ethanol dispersion of carbon black onto a carbon film (EMS CF300-Cu-UL Carbon Support Film), followed by drying in a dust-free environment.}
    \label{fig:figure0}
\end{figure}

Samples were prepared by dissolving a sodium salt of carboxymethylcellulose (NaCMC) from Sigma Aldrich (ref. 419303, with $M_w=250~\rm kg.mol^{-1}$ and $\mathrm{DS}=0.9$) in deionized water. The actual characteristics were determined to be $M_w=213~\rm kg.mol^{-1}$ and $\mathrm{DS}=0.88$ (see ref.~\cite{Legrand:2025} for details). NaCMC stock polymer solutions were prepared at various concentrations up to 5\% wt.~and stirred on a tube roller mixer, at room temperature, for 48~h until fully homogeneous. Carbon Black (CB) particles (VXC72R, Cabot Corp.) were then added to the solution. These hydrophobic colloidal particles consist of spherical-like primary particles of typical diameter $20~\rm nm$ irreversibly fused into tree-like, branched structures of typical diameter $500~\rm nm$ \cite{Richards:2017}. These particles display a broad diversity of shapes, as illustrated in Fig.~\ref{fig:figure0} where transmission electron micrographs (JEOL JEM-1400 TEM) of individual CB particles are displayed.   
  
The CB-CMC dispersions were sonicated in an ultrasonic bath for two sessions of $90~\rm min$ each, separated by 24~h of stirring with a magnetic mixer. After another 24~h of rest, the samples are ready for testing. Throughout the manuscript, samples are referred to by their CMC concentration $c_{\rm CMC}$ expressed in $\rm wt.\%$ of the stock solution (without CB) and their mass fraction of CB $x_{\rm CB}$, using the same notations as in ref.~\cite{Legrand:2023}.

\subsection{Rheometry}
\label{subsec:rheometry}

The rheological properties of CB-CMC hydrogels were measured using a cone-and-plate geometry (cone angle: 2$^\circ$, radius: $20~\rm mm$, and truncation gap: $46~\rm \mu$m) mounted on a strain-controlled rheometer (ARES G2, TA Instruments). The cone was smooth, while the bottom plate was sandblasted with a surface roughness of approximately 1~$\mu$m to prevent wall slip. All experiments were performed at constant temperature $T=22^\circ \rm C$, maintained thanks to a Peltier element located beneath the bottom plate. 

The rheological protocol consisted of three consecutive steps: ($i$) a preshear at a shear rate $\dot \gamma=500$~s$^{-1}$ for $3~\rm min$ to erase the loading history and rejuvenate the sample; ($ii$) a recovery phase of $20~\rm min$, during which the sample linear viscoelastic properties were monitored via small amplitude oscillations (strain amplitude $\gamma_0=0.03-0.1$\%) at a frequency of $\omega = 2\pi ~\rm rad.s^{-1}$; ($iii$) a strain sweep --hereafter referred to as LAOStrain test-- ranging from $\gamma_0=0.01\%$ to $300$\% or $3000\%$, depending on the hydrogel's conductive or insulating nature. The sweep was performed using 10 points per decade and 2 cycles per point, with only the last cycle used for measurement. All strain sweeps were conducted at  $\omega = 2\pi ~\rm rad.s^{-1}$ except for a series of experiments performed at various frequencies and discussed in Appendix~\ref{Appendix:ImpactFreq}.
LAOStrain data were analyzed within individual oscillation cycles, following the framework developed by Ewoldt \textit{et al.} \cite{Ewoldt:2008}. Raw stress and strain signals were recorded for each cycle using the transient acquisition mode of the Trios Software. To enable accurate frequency analysis and capture higher harmonics, a high sampling rate was used: for an input strain frequency of $2\pi~\rm rad.s^{-1}$, data were recorded at $256~\rm Hz$, allowing precise detection of harmonics up to the 9th order. A detailed discussion of the intracycle analysis is provided in Section~\ref{subsec:Intra}.

\subsection{Rheo-electric measurements}
\label{subsec:rheoelec} 

Rheo-electric measurements were performed using a stress-controlled rheometer (MCR 302, Anton Paar) equipped with a commercial Dielectric-Rheological Device (DRD, Anton Paar), previously employed for studying various types of complex fluids \cite{Erfanian:2021,Paulovicova:2025,Bauland:2025}. In practice, we used a $20~\rm mm$ radius parallel-plate geometry with a $1~\rm mm$ gap. The bottom plate was connected to a Peltier element for temperature control set at  $T=22^\circ \rm C$, while the upper plate was connected to the rheometer. Both plates also served as electrodes and were therefore chosen smooth, which may lead to some wall slip for non-linear deformations, as discussed in Section~\ref{sec:FTR}. Because the rheometer uses a feedback loop, each step of the strain sweeps is required to oscillate for about 5 cycles before reaching the desired value, and recording the last two cycles for analysis. 

The electrical circuit is closed using a loose copper wire clamped between two rings of a tightly coiled spring, which is attached to the rotor. The dimensions of the wire (total length $\sim 
5~\rm cm$, diameter $\sim 500~\rm \mu m$) were chosen to minimize its impact on mechanical measurements. To measure the electrical properties, a constant voltage $U_0 = 100~  \rm mV$ was applied using a potentiostat (SP-300, Biologic), and the resulting current $I$ was recorded as a function of time at a sampling rate of $100~\rm Hz$, much greater than the mechanical oscillation frequency $\omega = 2\pi~\rm rad.s^{-1}$. The electrical conductivity $\sigma_{\rm DC} $ was then computed via the cell constant $k$, using the relation $\sigma_{\rm DC} = k I / U_0$. The cell constant was determined to be $k = 0.013 \pm 0.005~\rm cm^{-1}$ based on calibration with reference KCl solutions \cite{legrand:tel-04768650}.

Finally, the finite rigidity of the copper wire attached to the rotor may affect the rheological measurements, and its impact was estimated to be, at most, of about $5~\rm Pa$ over the range of strain values explored during the LAOStrain test (see Appendix~\ref{appendix:rheoelec1}). We, therefore, chose to study samples with viscoelastic properties much larger than $5~\rm Pa$ for rheo-electrical tests. In addition, the displacement of the wire during the LOAStrain test also impacts the electrical measurements, yielding current fluctuations, yet uncorrelated with the strain amplitude (see discussion in Appendix~\ref{appendix:rheoelec2}).

\begin{figure*}[!th]
    \centering
\includegraphics[width=0.7\linewidth]{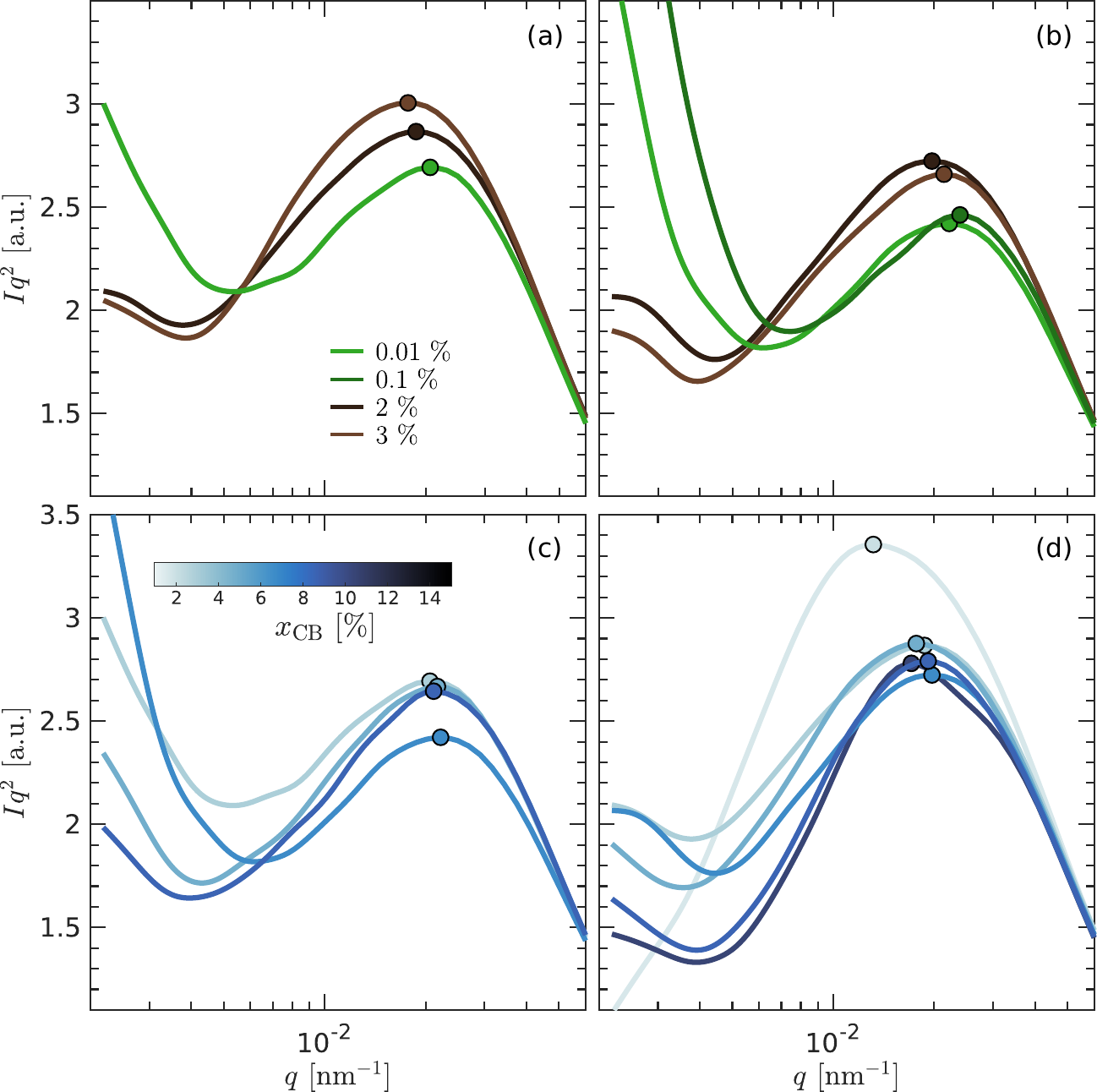}
    \caption{Kratky plots of the scattering intensity ($Iq^2$ vs.~$q$) for CB-CMC dispersions of varying compositions. (a) and (b) Fixed CB content at $x_{\rm CB} =2\%$ and $x_{\rm CB} =6\%$, respectively, with CMC concentrations $c_{\rm CMC} = 0.01, 0.1, 2,$ and $3\%$. Electrically conductive hydrogels ($r<r_c$) are shown in green; insulating hydrogels ($r>r_c$) in brown. (c) and (d) Fixed CMC concentration at $c_{\rm CMC} = 0.01\%$ and $c_{\rm CMC} = 2\%$, respectively, with CB content $x_{\rm CB}$ ranging from $1\%$ to $10\%$. Darker colors indicate higher CB content, as shown by the color bar in (c). In all figures, disk symbols mark the local maxima of the Kratky plots. All data have been normalized to collapse at large $q$-values to account for variations in CB content.}
    \label{fig:figure1}
\end{figure*}

\subsection{USAXS and Rheo-TRUSAXS measurements}
\label{subsec:rheoUSAXS} 

Ultra-Small-Angle X-ray Scattering (USAXS) experiments were performed at the ID02 beamline of the ESRF (Grenoble, France) using a monochromatic, highly collimated, and intense X-ray beam with a wavelength of $1.013$~\AA~  ($12.230~\rm kV$). Experiments were conducted at a sample-to-detector distance of $30.7~\rm m$, allowing access to ultra-small angles and covering a scattering vector range of $2\times 10^{-3} \leq q \leq 10^{-1}~\rm nm^{-1}$. Two-dimensional (2D) SAXS patterns were recorded on an Eiger2-4M pixel array detector and collapsed at large $q$-values after appropriate background subtraction and transmission normalization (see Ref.~\cite{Narayanan:2022} for technical details). For static experiments, we used $2~\rm mm$ diameter borosilicate capillaries (WJM-Glas M\"uller GmbH).

\begin{figure*}[!t]
    \centering
    \includegraphics[width=0.9\linewidth]{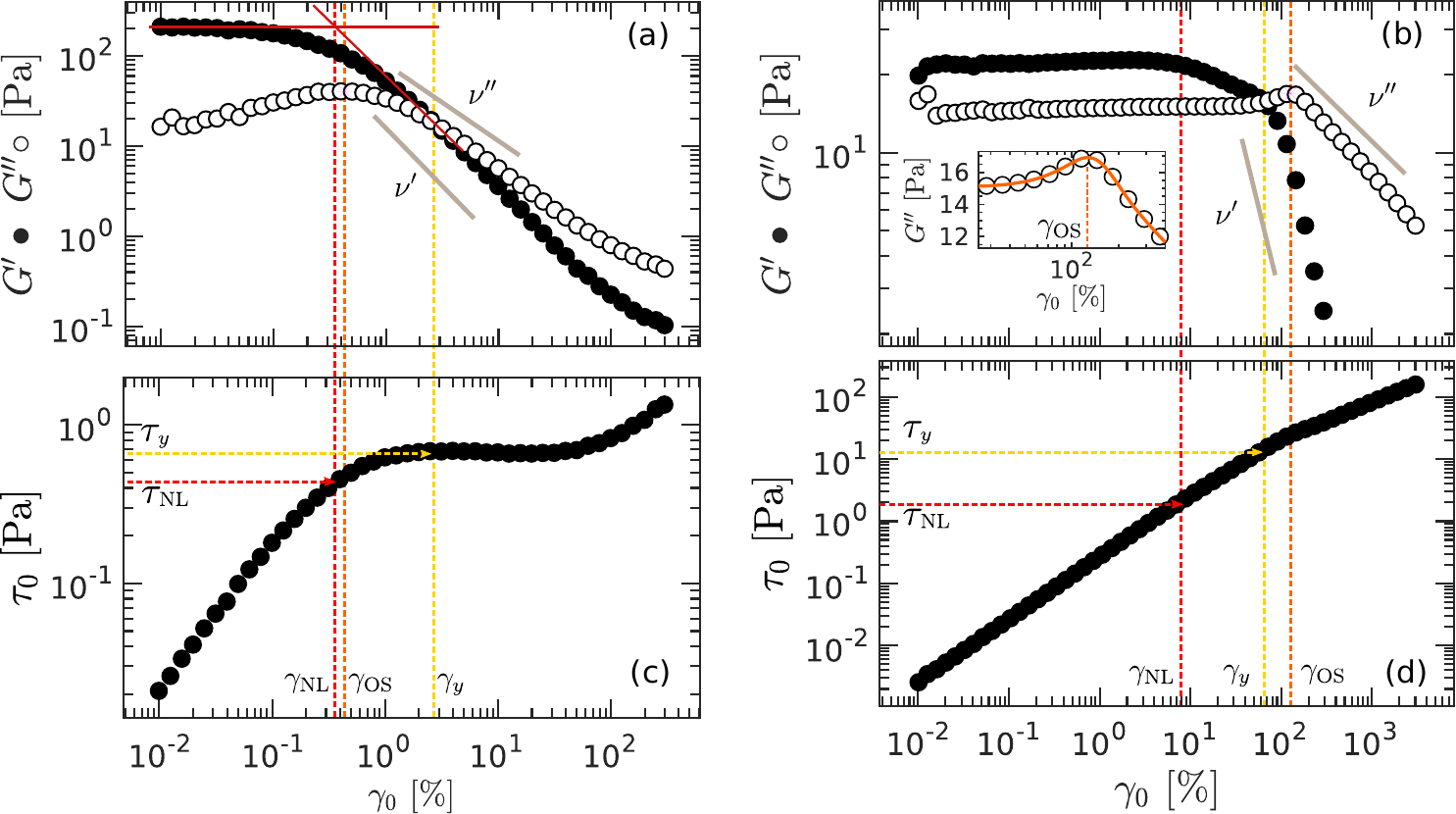}
    \caption{LAOStrain response of conductive (left) and insulating (right) CB-CMC hydrogels. Elastic modulus $G'$ (full symbols)  and viscous modulus $G''$ (empty symbols) vs.~strain amplitude $\gamma_0$ measured by oscillatory shear at $\omega = 2\pi ~\rm rad.s^{-1}$ for (a) a conductive CB-CMC hydrogel ($c_{\rm CMC} = 0.01\%$ and $x_{\rm CB} = 6\%$, $r<r_c$), and (b) an insulating CB-CMC hydrogel ($c_{\rm CMC} = 2\%$ and $x_{\rm CB} = 8\%$, $r>r_c$). The corresponding stress amplitude $\tau_0(\gamma_0)$ is plotted in (c) and (d). The vertical dashed lines indicate the strain $\gamma_{\rm NL}$ marking the onset of non-linearity (red), the strain $\gamma_{\rm OS}$ at which $G''$ is maximum (orange), and the yield strain $\gamma_y$ defined by the crossover of $G'$ and $G''$ (yellow). The corresponding stress values $\tau_{\rm NL}$ and $\tau_y$ are shown as horizontal dashed lines in (c) and (d). For both types of hydrogels, $G'$ and $G''$ decrease as power-laws of $\gamma_0$, with exponents $\nu'$= 0.92 and $\nu''$=0.65 in (a), and $\nu'=$0.71 and $\nu''=$0.34 in (b) shown as gray lines. The red solid lines in (a) show the geometrical construction to obtain $\gamma_{\rm NL}$ as the intersection between the extrapolation of the linear regime (horizontal solid line) and the tangent of $G^\prime$ at $\gamma_0=\gamma_y$.  The inset in (b) illustrates the determination of $\gamma_{\rm OS}$ using a smoothed and extrapolated version of $G^{\prime\prime}$ vs $\gamma_0$, with a spline method represented by the orange curve.}
    \label{fig:strain_sweep_CB8}
\end{figure*}

USAXS measurements were coupled with rheometry to perform Time-Resolved USAXS (TRUSAXS) and characterize the microstructure of CB-CMC hydrogels under shear. Samples were loaded into a polycarbonate Couette cell, composed of a bob with a radius of $10~\rm mm$, and a cup with a radius of $11~\rm mm$, and a height of $40~\rm mm$, connected to a stress-controlled rheometer (Haake RS6000, Thermo Scientific). The X-ray beam was aligned in the so-called ``radial" configuration, passing across the Couette cell along the direction of the velocity gradient $\nabla \vec{v}$ -- i.e., through the rotation axis of the bob. Scattering patterns were thus recorded in the (velocity $\vec{v}$, vorticity $\nabla \times \vec{v}$) plane. 

Because scattering intensity scales with the square of the difference in scattering length density (SLD) between phases, and given that CB has an SLD of $3.1 \times 10^{11}\rm cm^{-2}$ compared to $9.54 \times 10^{10}\rm cm^{-2}$ for water and $5.5 \times 10^{10}~\rm cm^{-2}$ for CMC, the contribution of CB to the total scattered intensity is approximately 20 times larger than that of the CMC matrix. As a result, the scattered intensity of CB-CMC hydrogels is dominated by the CB particles in the explored $q$ range.
To isolate the signal arising from the CB microstructure, the background scattering was systematically subtracted: both the intensity scattered by the Couette cell filled with distilled water, and that of a pure CMC suspension at $c_{\rm CMC}=2\%$  were measured and subtracted from the intensity of the CB-CMC hydrogels, taking into account the actual CMC concentration. This rheo-TRUSAXS setup allows us to perform TRUSAXS measurements under large amplitude oscillatory shear, as analyzed in detail in Section~\ref{sec:RheoTRUSAXS}.

\section{Microstructural properties of CB-CMC hydrogels}
\label{sec:Results0} 

In this section, we present a quantitative analysis of the microstructure of CB-CMC hydrogels based on static USAXS measurements performed on CB-CMC hydrogels spanning both the conductive and insulating regimes. The results are reported as Kratky plots, $Iq^2$ vs $q$, in Figure~\ref{fig:figure1}.

The impact of CMC concentration at fixed CB content is illustrated in Figs.~\ref{fig:figure1}(a) and \ref{fig:figure1}(b) for $x_{\rm CB}=2\%$ and $6\%$, respectively. All compositions yield Kratky plots with a characteristic third-order polynomial shape featuring a local maximum around $q_{\rm max} \simeq 2.10^{-2}~\rm nm^{-1}$, corresponding to a real-space length scale $d_{\rm max}=2\pi/q_{\rm max} \simeq 300~\rm nm$. In systems comprising attractive scatterers, this seemingly unique length-scale actually reflects the interplay between two distinct contributions: (i) the form factor, whose shoulder in the $I(q)$ representation gives rise to the observed peak in the Kratky one, corresponding here to the diameter of the CB aggregates, and (ii) the structure factor, whose shape is linked to the most probable center-to-center distance between closest neighbors. For monodisperse attractive hard spheres, these two length scales coincide, but in more complex or polydisperse systems, they are partially decoupled \cite{Lindner:1991}. Accordingly, relative variations of $q_{\rm max}$ can be interpreted in terms of modifications to either or both of these features. Here, as the CMC concentration increases, the hydrogel transitions from conductive (green curves) to insulating (brown curves) \cite{Legrand:2023}, and the local maximum shifts towards lower $q$ values. This shift is indicative of an increasing separation between CB aggregates, consistent with a progressive loss of connectivity and with the onset of a depercolation transition \cite{baeza2015depercolation} as the polymer-to-particle ratio $r$ crosses the critical threshold $r_c$. In parallel, the peak intensity at $q_{\rm max}$ increases with CMC concentration, which indicates a higher degree of order of the CB spatial distribution, likely due to a stronger steric repulsion caused by the CMC adsorption onto the CB particles \cite{baeza2013effect, baeza2016revealing}.
Beyond the position and intensity of the local maximum, the Kratky plots in Figs.~\ref{fig:figure1}(a) and \ref{fig:figure1}(b) reveal notable differences in the low-$q$ range ($q<0.01~\rm nm^{-1}$), where $Iq^2$ exhibits a local minimum. Conductive CB-CMC hydrogels display a pronounced upturn at low $q$, indicative of long-range correlations in the CB spatial distribution, consistent with a percolated CB network. In contrast, insulating CB-CMC hydrogels exhibit only a weak increase in this low-$q$ range, confirming that CB particles form, at best, only small, isolated clusters embedded in the CMC polymer matrix.

The impact of the CB content at fixed CMC concentration is presented in Figs.~\ref{fig:figure1}(c) and \ref{fig:figure1}(d) for $c_{\rm CMC}=0.01\%$ and $c_{\rm CMC}=2\%$, respectively. These results can be interpreted in light of prior work by Baeza \textit{et al}. \cite{baeza2013multiscale} and Genix \textit{et al}. \cite{genix2017determination} on aggregated particulate systems, highlighting two key features. First, when $c_{\rm CMC}=0.01\%$, increasing the CB content does not have an effect on $q_{\rm max}$,  indicating that, in the absence of significant steric repulsion, CB particles form large agglomerates in which the most probable distance between neighboring particles corresponds to their diameter, close to $300~\rm nm$. A similar trend is observed at $c_{\rm CMC}=2\%$, except at the lowest CB content ($x_{\rm CB}=1\%$), where a decrease in $q_{\rm max}$ suggests that the CMC concentration is sufficient to repel the CB particles from each other. In line with the above-mentioned arguments, it is also worth noting that the peak intensity is larger in Fig.~\ref{fig:figure1}(d) than in Fig.~\ref{fig:figure1}(c), reflecting stronger repulsive interactions due to increased CMC adsorption. Second, the intensity of the $q_{\rm max}$ peak appears largely insensitive to the CB content, despite a tenfold variation in particle concentration, in both Figs.~\ref{fig:figure1}(c) and \ref{fig:figure1}(d). Although counterintuitive, this behavior can be rationalized by the high polydispersity in the size of CB particles --estimated to exceed 40\% \cite{Donnet:1993}-- which hinders the formation of coherent interparticle correlations, thus suppressing constructive interferences. However, a clear trend is observed in the lower-$q$ region: as the CB content increases, the depth of the so-called ``correlation-hole'' also increases, corresponding to a reduction in scattering intensity and reflecting a lower isothermal compressibility of the material \cite{baeza2013multiscale}. 
In addition, the upturn of intensity at very low $q$ is not affected by the CB content in both Figs.~\ref{fig:figure1}(c) and \ref{fig:figure1}(d). In Fig.~\ref{fig:figure1}(c), the steep intensity upturn suggests the presence of a branched network, whereas the more moderate increase in Fig.~\ref{fig:figure1}(d) is indicative of the presence of more isolated aggregates. 
In summary, the observed difference in scattering intensity between Figs.~\ref{fig:figure1}(c) and \ref{fig:figure1}(d) is primarily attributed to changes in the spatial distribution of particles, i.e., related to the structure factor.
These results confirm the picture originally proposed in our previous work \cite{Legrand:2023}, in which increasing the polymer concentration at fixed CB content results in a \textit{depercolation transition}, i.e., a dispersion of the CB particles in the form of isolated clusters within the CMC polymer matrix. 

In the rest of the text, we examine the non-linear rheological response of CB-CMC hydrogels and compare the response of both types of gels to strain-controlled Large Amplitude Oscillatory Shear tests.

\section{\label{sec:Results1} LAOStrain response of CB-CMC hydrogels}

\subsection{\label{subsec:Compo} Impact of CB-CMC hydrogels composition on the key non-linear features of the LAOStrain response}

We now turn to the response of CB-CMC hydrogels under oscillatory shear of increasing amplitude $\gamma_0$. We shall emphasize that all experiments are conducted at a fixed frequency ($\omega = 2\pi~\rm rad.s^{-1}$), although conductive and isolated CB-CMC hydrogels exhibit different frequency-dependence in the linear regime --namely, almost frequency independent for conductive gels, and power-law for insulating gels \cite{Legrand:2023}. Representative LAOStrain results for both a conductive and an insulating CB-CMC hydrogel are shown in Figure~\ref{fig:strain_sweep_CB8}. At low strain amplitudes, both types of hydrogels show a solid-like response: the elastic modulus $G'$ remains larger than the viscous modulus $G''$ until the onset of non-linearity, which is reached at $\gamma_0=\gamma_{\rm NL}$, whose definition is based on a geometrical construction (see caption of Fig.~\ref{fig:strain_sweep_CB8}). Beyond this point, for larger strain amplitudes, CB-CMC hydrogels undergo a so-called type III yielding transition \cite{Hyun:2002}, characterized by a continuous decrease in $G'$, and an overshoot in $G''$, peaking at $\gamma_0=\gamma_{\rm OS}$. This overshoot in $G''$ is reminiscent of the ``\textit{Payne effect}'' observed in filler-polymer composites \cite{Xu:2019,Fan:2019, shi2021influence}, which reflects an excess of dissipation during the yielding process that the steric stabilization of the fillers can reduce \cite{hart2025structural}, following a progressive and heterogeneous failure scenario \cite{Donley:2020}.  Finally, the yield point is defined as the crossover of $G'$ and $G''$, which occurs at $\gamma_0=\gamma_y$.

\begin{figure}[!t]
    \centering
    \includegraphics[width=0.9\linewidth]{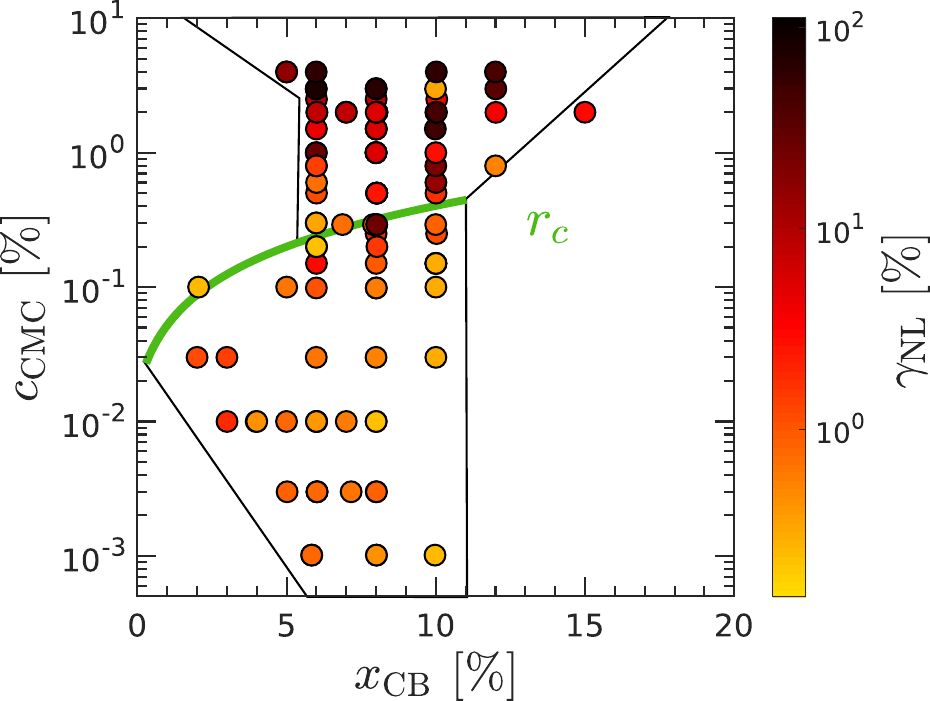}
    \caption{Phase diagram of aqueous CB-CMC dispersions as a function of CB weight fraction $x_{\rm CB}$ and CMC weight fraction $c_{\rm CMC}$. The black lines delineate the gel region, as identified in Ref.~\cite {Legrand:2023} through linear viscoelastic measurements. Color levels code for the strain $\gamma_{\rm NL}$ marking the onset of non-linearity. The green curve corresponds to the critical polymer-to-particle ratio $r=r_c$, determined from linear viscoelastic and electrical impedance spectroscopy measurements \cite{Legrand:2023}, separating conductive ($r<r_c$) and insulating ($r>r_c$) hydrogels.}
    \label{fig:figure3}
\end{figure}

\begin{figure*}[!th]
    \centering
    \includegraphics[width=0.9\linewidth]{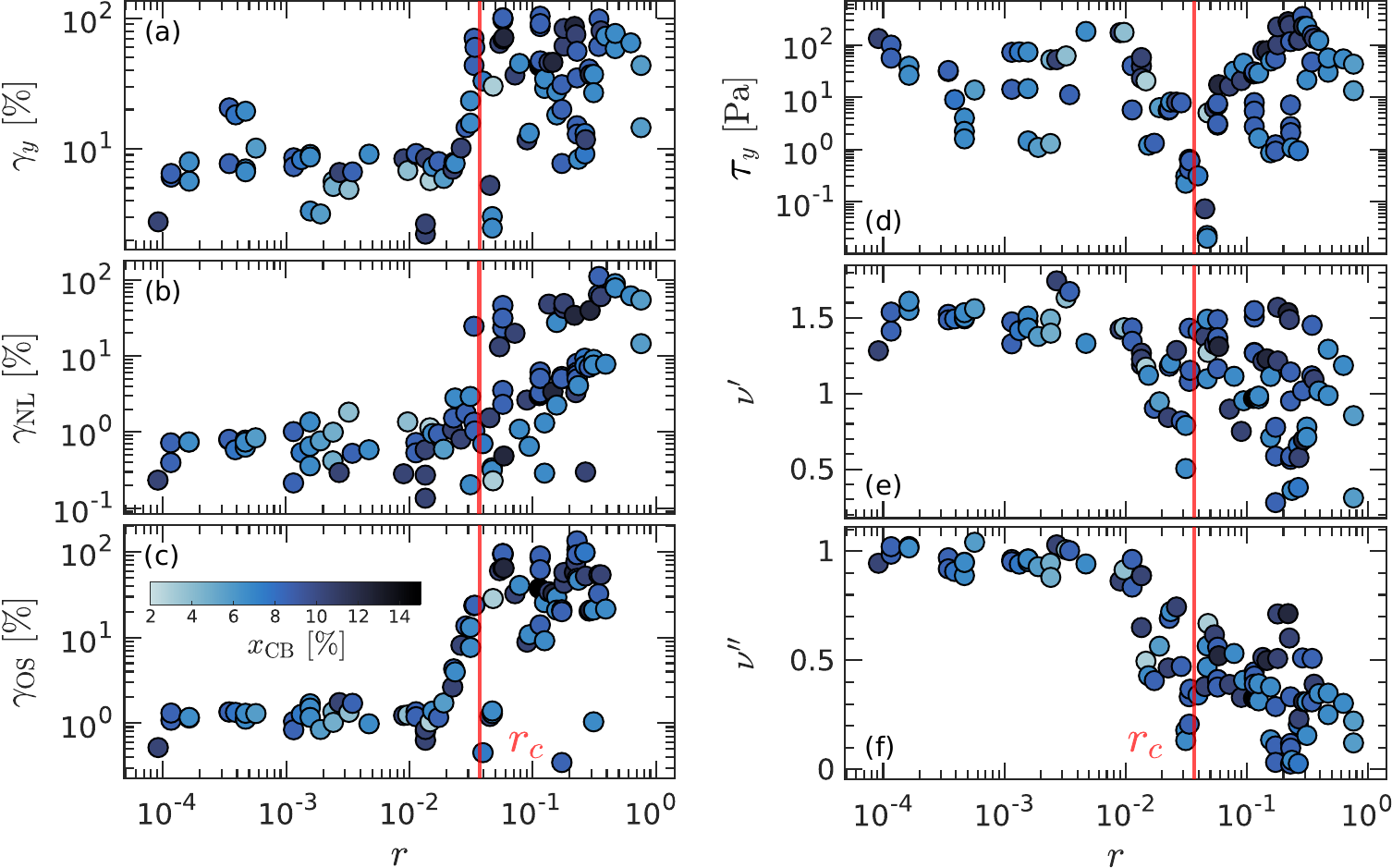}
    \caption{Attributes of the non-linear response and yielding transition of CB-CMC hydrogels plotted against the CMC-to-CB ratio $r$. Three characteristic strain values are reported: (a) yield strain $\gamma_y$, (b) onset of the non-linear regime $\gamma_{\rm NL}$, and (c) the strain $\gamma_{\rm OS}$ at which $G''$ is maximum. The yield stress $\tau_y = \tau_0(\gamma_y)$ is reported in (d), while the exponents $\nu'$ and $\nu''$ characterizing the power-law decay of $G'$ and $G''$ beyond the yield point are reported in (e) and (f), respectively. Colors encode for the CB content; darker colors correspond to larger $x_{\rm CB}$ [see color bar in (c)]. In all graphs, the vertical red line represents the critical ratio $r_c \simeq 0.037$ separating conductive from insulating hydrogels, as determined by previous linear viscoelastic and electrical impedance spectroscopy measurements \cite{Legrand:2023}.}
    \label{fig:strain_sweep_analyzed}
\end{figure*}

Despite such similarities, the non-linear responses of conductive and insulating CB-CMC hydrogels differ by several key factors. First, the conductive CB-CMC hydrogel yields at $\gamma_y\simeq 3\%$, a value typical of colloidal gels including colloid-polymer depletion gels with non-adsorbing polymers \cite{Muller:2023,Laurati:2014}, whereas the insulating CB-CMC hydrogel yields at a much larger strain, $\gamma_y\simeq 60\%$. Such a discrepancy supports the interpretation that the viscoelastic properties of conductive hydrogels are governed by a percolated network of CB particles, while that of insulating hydrogels is dominated by the CMC polymer matrix \cite{Legrand:2023}.
Second, the overshoot in $G''$ is significantly more pronounced in the conductive gel, reaching an amplitude nearly 3 times its linear viscoelastic value. In contrast, the insulating CB-CMC hydrogel shows only a modest overshoot, with $G''$ increasing by a few percent. In the latter case, the amplitude of the overshoot strongly depends on the frequency of the oscillations, and goes to zero for increasing frequency [see discussion and Fig.~\ref{fig:CB_CMC_non_lin_strain_sweep_different_frequencies} in Appendix~\ref{Appendix:ImpactFreq}]. 
Furthermore, the location of the overshoot relative to the yield point differs: in the conductive hydrogel, the overshoot precedes the yield point (i.e., $\gamma_{\rm OS}<\gamma_y$), whereas in the insulating hydrogel, it occurs beyond the yield point ($\gamma_{\rm OS}>\gamma_y$). 
Third, the post-yield response of the hydrogels is characterized by a power-law decay of $G'$ and $G''$ for increasing strains, i.e., $G' \sim \gamma_0^{-\nu'}$ and $G'' \sim \gamma_0^{-\nu''}$ where the exponents $\nu'$ and $\nu''$ depend on the nature of the hydrogel: $\nu'=0.92$ and $\nu''=0.65$ in the case of the conductive gel [Fig.~\ref{fig:strain_sweep_CB8}(a)], and  $\nu'=0.71$ and $\nu''=0.34$ for the insulating gel [Fig.~\ref{fig:strain_sweep_CB8}(b)]. This also leads to a strong difference in the ratio of $\nu''$ and $\nu'$, with $\nu''/\nu' =1.4$ for the conductive gel and $\nu''/\nu' = 2.1$ for the insulating gel, which suggests that the yielding scenario differs in these two types of gels, as further discussed below.

\begin{figure*}[!th]
    \centering
    \includegraphics[width=0.7\linewidth]{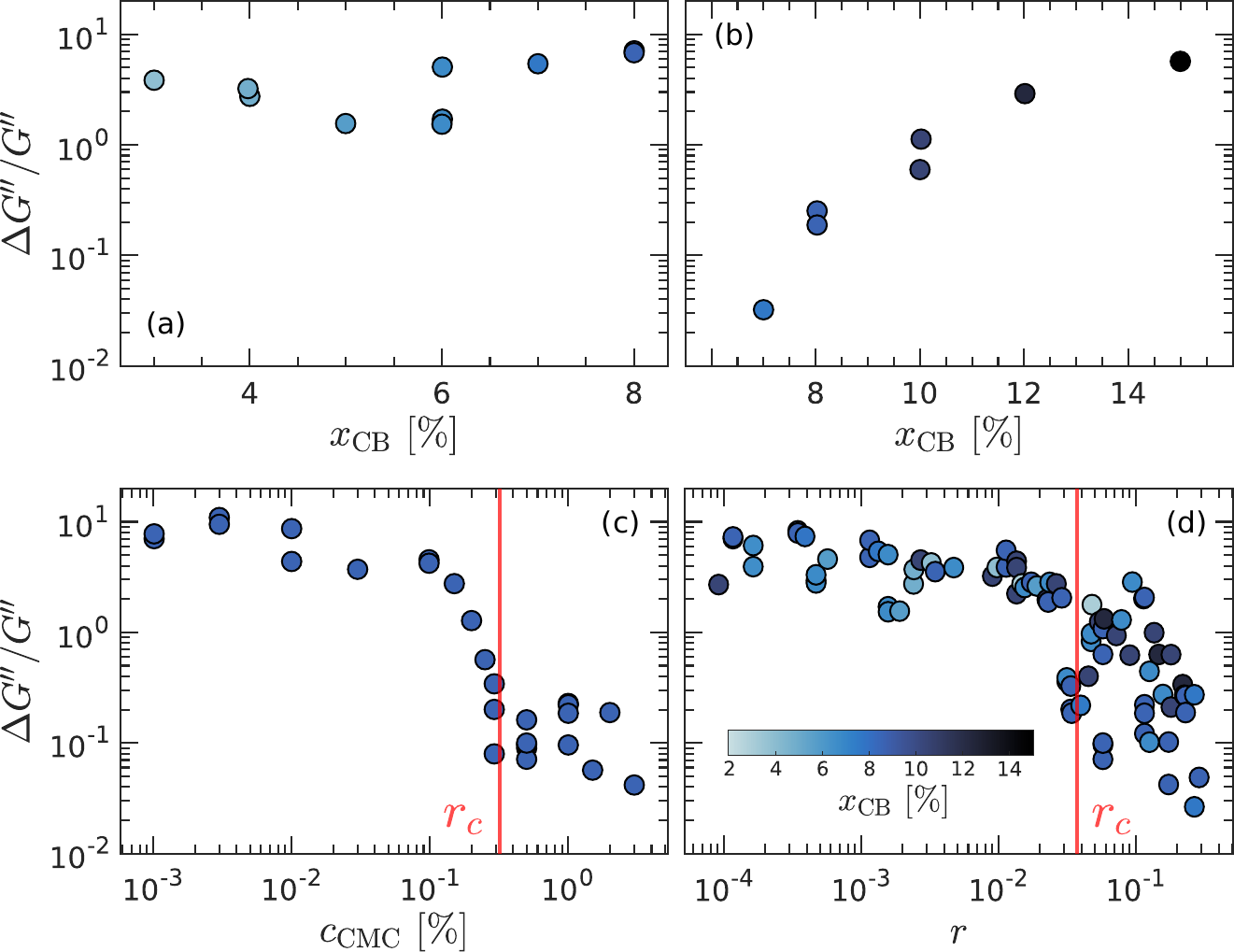}
    \caption{Key features of the overshoot in $G''$ observed during LAOStrain tests on CB-CMC hydrogels. Impact of the CB content $x_{\rm CB}$ on the relative amplitude of the overshoot in $G''$ for a fixed CMC concentration in (a) a conductive CB-CMC hydrogel $c_{\rm CMC} = 0.01\%$ and (b) an insulating CB-CMC hydrogel $c_{\rm CMC} = 2\%$. (c) Impact of the CMC concentration on the relative amplitude of the overshoot in $G''$ for a fixed CB content $x_{\rm CB} = 8\%$. (d) Relative amplitude of the overshoot in $G''$ as a function of the CMC-to-CB ratio $r$ for all compositions tested. Color codes for the CB content. In (c) and (d), the vertical red line indicates $r = r_c$, which separates electrically conductive from insulating hydrogels, as identified by previous linear viscoelastic measurements \cite{Legrand:2023}.}
    \label{fig:overshoot}
\end{figure*}

These two markedly different yielding behaviors are consistently observed across a broad range of compositions. To confirm their robustness, we prepared and tested over one hundred CB-CMC hydrogels spanning a wide range of formulations with $0.001\%\leq c_{\rm CMC}\leq$4\% and 2\%$\leq x_{\rm CB}\leq$15\%. As a representative example,  Fig.~\ref{fig:figure3} shows a phase diagram mapping the strain at the onset of non-linearity $\gamma_{\rm NL}$ as a function of CMC concentration and  CB content. This phase diagram can be directly compared to the one constructed from linear viscoelastic measurements and previously reported in ref.~\cite{Legrand:2023}. The values of $\gamma_{\rm NL}$ naturally form two different regions: lower values at low CMC concentrations and higher values at high CMC concentrations. These two regions coincide with the domains attributed to the conductive and insulating hydrogels, respectively, which are separated by the critical CMC-to-CB ratio $r_c$, represented by the green curve in Fig.~\ref{fig:figure3}. A similar conclusion is reached when constructing a phase diagram based on the yield strain $\gamma_y$ instead of $\gamma_{\rm NL}$ \cite{legrand:tel-04768650}.

\begin{figure*}[!th]
    \centering
    \includegraphics[width=0.6\linewidth]{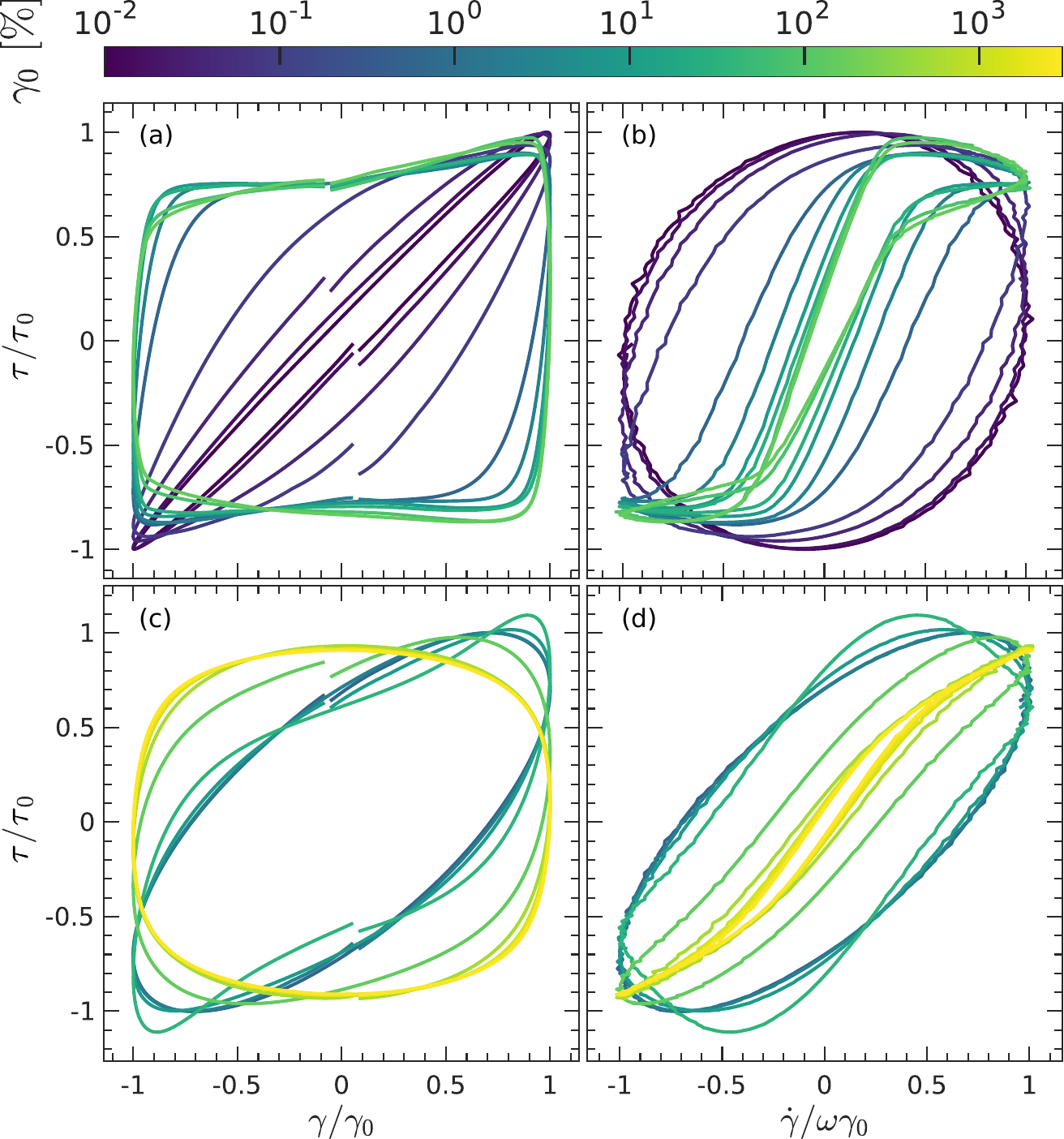}
    \caption{Lissajous-Bowditch plots of the normalized stress response under sinusoidal strain of increasing amplitude. Experiments were conducted at $\omega = 2\pi~\rm rad.s^{-1}$ on two types of CB-CMC hydrogels: (a,b) an electrically conductive sample  ($c_{\rm CMC} = 0.01\%$ and $x_{\rm CB} = 6\%$, $r<r_c$), and (c,d) an insulating sample ($c_{\rm CMC} = 3\%$ and $x_{\rm CB} = 8\%$, $r>r_c$). The left column shows the instantaneous stress $\tau$ as a function of the instantaneous strain $\gamma$, while the right column shows the same stress data plotted against the instantaneous shear rate $\dot\gamma$. Each observable $\tau$, $\gamma$, and $\dot\gamma$ is normalized by the amplitude of its fundamental harmonic at frequency $\omega$, respectively $\tau_0$, $\gamma_0$, and $\omega\gamma_0$. Colors encode the strain amplitude $\gamma_0$ (see color bar at the top of the figure).}   
    \label{fig:Lissajous}
\end{figure*}

More quantitatively, the key features of the LAOStrain response of CB-CMC hydrogels, namely the onset of non-linearity $\gamma_{\rm NL}$, the yield point coordinate ($\gamma_y$, $\tau_y$), the strain at which the overshoot in $G''$ is maximum ($\gamma_{\rm OS}$), and the power-law exponents $\nu'$ and $\nu''$ are reported for all investigated compositions as a function of the polymer-to-particle ratio $r$ in Fig.~\ref{fig:strain_sweep_analyzed}. 
The critical ratio $r_c$, identified previously from linear viscoelastic and electrical impedance spectroscopy measurements as the depercolation threshold \cite{Legrand:2023}, is indicated as a red vertical line. Strikingly, the critical ratio $r_c$ delineates a sharp transition between two distinct non-linear rheological responses, particularly evident for  $\gamma_{\rm NL}$, $\gamma_{\rm OS}$, and $\gamma_y$ [Figs.~\ref{fig:strain_sweep_analyzed}(a)-(c)]. These observables highlight the contrast between the non-linear response of conductive ($r<r_c$) and insulating ($r>r_c$) CB-CMC hydrogels. 
In contrast, the yield stress $\tau_y$ shows similar values on both sides of $r_c$, albeit with a large drop at $r=r_c$, where the linear elastic properties of the hydrogels are the weakest due to the depercolation transition \cite{Legrand:2023}. 
Moreover, the exponents $\nu'$ and $\nu''$, which characterize the power-law decay of $G'$ and $G''$ beyond the yield point, also differ markedly across the transition. For conductive hydrogels ($r<r_c$), we observe that $\nu'\simeq 1.5$ and $\nu'' \simeq 1$, and therefore that $\nu'/\nu'' \simeq 1.5$ (see also Fig.~\ref{fig:ratio_nu} in Appendix~\ref{Appendix:rationu}). This value is slightly smaller than the prediction by mode coupling theory ($\nu'/\nu'' =2$), which applies to yield stress fluids with a simple Maxwell-like behavior in the vicinity of the yield point \cite{Miyazaki:2006,Wyss:2007}, and comparable to that encountered in colloidal glasses, where $\nu'/\nu''\lesssim  2$, at least for sufficiently high volume fractions \cite{Koumakis:2012,Helgeson:2007}. Interestingly, such a comparison is in line with the glassy-like viscoelastic spectrum of conductive CB-CMC hydrogels measured in the linear deformation regime \cite{Legrand:2023}.
As $r$ increases and crosses $r_c$, $\nu'$ spans a broad range, from $0.5$ to $1.5$, without any clear trend, while $\nu''$ decreases monotonically from $1$ to about $0.3$ [Figs.~\ref{fig:strain_sweep_analyzed}(e)-(f)]. As a result, for insulating gels, the ratio $\nu'/\nu''$ increases with $r$  ($r>r_c$), reaching values as high as $\nu'/\nu'' \simeq 5$  (see also Fig.~\ref{fig:ratio_nu} in Appendix~\ref{Appendix:rationu}). Finally, the fact that $\nu'/\nu'' \gg 2$ suggests that insulating CB-CMC hydrogels are not well described by a Maxwell model in the vicinity of the yield point \cite{Miyazaki:2006}, but instead must be characterized by a broader relaxation spectrum, which is already the case in the linear deformation regime \cite{Legrand:2023}.  
Notably, the exponents $\nu'$ and $\nu''$ display greater variability in insulating hydrogels. This may reflect a more heterogeneous microstructure comprising isolated clusters of CB particles with a broad size distribution, and whose exact distribution may depend on shear history and sample preparation. In addition, the variability of $\nu'$ and $\nu''$ is likely impacted by the diversity of polymer segments (including polymer entanglement, loops, and dangling ends as well as bridges between CB particles) and their length distribution.

Focusing now on the overshoot in $G''$, clearly visible in Figs.~\ref{fig:strain_sweep_CB8}(a) and \ref{fig:strain_sweep_CB8}(b), we observe that the location $\gamma_{\rm OS}$ of its maximum undergoes an abrupt transition across $r_c$: from $\gamma_{\rm OS}\simeq 1\%$ in conductive hydrogels ($r<r_c$) to $\gamma_{\rm OS}\simeq 50\%$ in insulating ones ($r>r_c$) [Fig.~\ref{fig:strain_sweep_analyzed}(c)]. Interestingly, $\gamma_{\rm OS}$ starts increasing already at $r \simeq 10^{-2} < r_c$, while the hydrogel remains conductive. It is only beyond the conductive-to-insulating transition that $\gamma_{\rm OS}$ levels off to a plateau value.  
Similarly, the relative amplitude of the overshoot, $\Delta G''/G''$, also serves as a strong indicator of microstructural changes across $r_c$. Indeed, in conductive CB-CMC hydrogels, the maximum in $G''$ at the overshoot is about five times larger than the linear $G''$ value, and remains largely independent of both CMC concentration and CB content, at least for $r \ll r_c$, i.e., sufficiently far from the conductive-to-insulating transition [Figs.~\ref{fig:overshoot}(a) and \ref{fig:overshoot}(c)]. In contrast, in the insulating regime, the overshoot is much weaker and its relative amplitude strongly increases for increasing CB content [Fig.~\ref{fig:overshoot}(b)], while being poorly sensitive to CMC concentration [Fig.~\ref{fig:overshoot}(c)]. Since the CB particles play the role of physical crosslinkers in the CMC matrix, this suggests that the amplitude of the overshoot is directly linked to the average number of adsorbed CMC polymer per CB particle present in the system at rest.

To summarize, these results can be interpreted as follows: in conductive gels, the overshoot in $G''$  originates from the percolated network of CB particles. Approaching the depercolation transition as $r \rightarrow r_c$ while $r \lesssim r_c$, the overshoot amplitude decreases continuously as the network of CB particles is progressively disrupted due to the relative increase in CMC content [Fig.~\ref{fig:overshoot}(d)]. Eventually, the overshoot is weaker but still exists in the insulating state due to the CB particles acting as physical crosslinkers in the CMC polymer matrix. In the latter case, the relative amplitude of the overshoot shows a greater variability. This is again likely due to the heterogeneity of the microstructure in insulating hydrogels, where the degree of dispersion of the CB particles affects both the number and the strength of the crosslinkers \cite{Eckman:2025}, which in turn, impacts the magnitude of the overshoot in $G''$.  

\subsection{\label{subsec:Intra} Analysis of LAOStrain ``intra-cycle'' waveforms}

\subsubsection{Lissajous-Bowditch (LB) curves}
\label{sec:LBcurves}

We now go beyond the cycle-averaged rheological response of CB-CMC hydrogels to oscillatory shear, and examine the intra-cycle stress response for increasing strain amplitudes. The results are reported in Fig.~\ref{fig:Lissajous} as normalized Lissajous-Bowditch (LB), i.e., parametric plots of the shear stress $\sigma$ as a function of strain $\gamma$ [Figs.~\ref{fig:Lissajous}(a) and \ref{fig:Lissajous}(c)] or strain rate $\dot \gamma$ [Figs.~\ref{fig:Lissajous}(b) and \ref{fig:Lissajous}(d)] at various values of the strain amplitude $\gamma_0$ along the LAOStrain test.

For the conductive CB-CMC hydrogel [Figs.~\ref{fig:Lissajous}(a)–(b)], at low strain amplitudes ($\gamma_0 \lesssim 0.1\%$), the $\sigma(\gamma)$ LB curves [Fig.~\ref{fig:Lissajous}(a)] are nearly elliptical, and the corresponding $\sigma(\dot \gamma)$ LB curves [Fig.~\ref{fig:Lissajous}(b)] are close to circular, indicative of a predominantly elastic, gel-like behavior with moderate energy dissipation. As the strain amplitude increases beyond $\gamma_0 \simeq 1\%$, the material enters a non-linear regime marked by pronounced deviations from the elliptical shape. The $\sigma(\gamma)$ LB curves become increasingly asymmetric and skewed toward the horizontal axis [Fig.~\ref{fig:Lissajous}(a)], while the $\sigma(\dot \gamma)$ LB curves narrow significantly and develop a distinct S-shape indicating that the stress becomes relatively insensitive to the rate of deformation for $\dot \gamma / \omega \gamma_0 \gtrsim 0.3$ [Fig.~\ref{fig:Lissajous}(b)]. Based on numerical results obtained with the SGR model \cite{Radhakrishnan:2018}, this plateau-like behavior suggests a localized yielding, where the material undergoes abrupt structural rearrangements that limit the buildup of internal stress. Such a feature is typical of soft materials with fragile, percolated networks, such as CB gels, where yielding under LAOS is spatially heterogeneous and governed by the rupture of stress-bearing paths within the particulate network \cite{Perge:2014}. 

\begin{figure*}[!ht]
    \centering
      \includegraphics[width=0.9\linewidth]{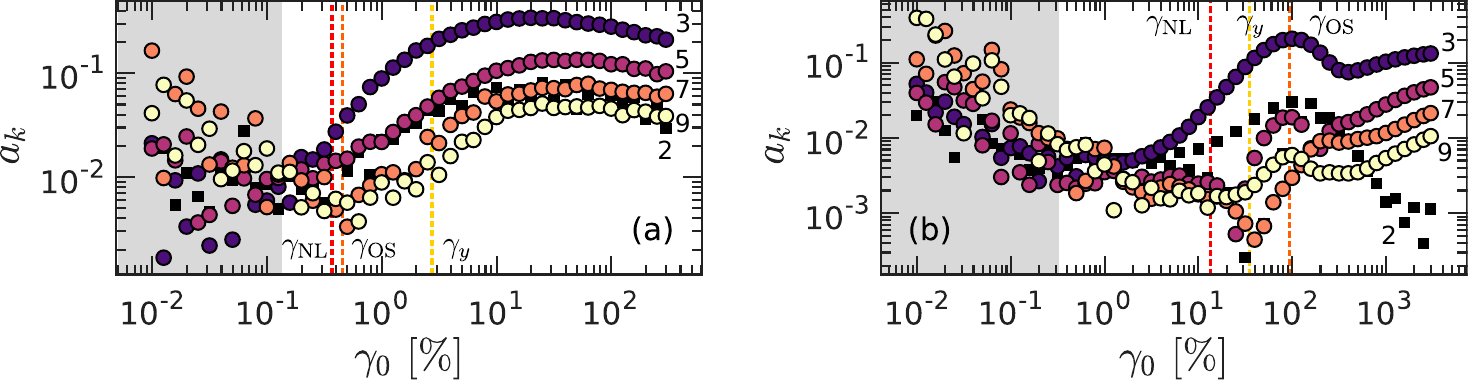}
    \caption{Normalized Fourier coefficients $a_k$ of the stress decomposition as a function of the strain amplitude $\gamma_0$ for the two types of CB-CMC hydrogels: (a) a conductive gel ($c_{\rm CMC} = 0.01\%$ and $x_{\rm CB} = 6\%$) and (b) an insulating gel ($c_{\rm CMC} = 2\%$ and $x_{\rm CB} = 8\%$). The orders $k=2, 3, 5, 7$, and 9 of the various harmonics are reported on the graph, with lighter colors corresponding to increasing values of $k$. The value $k=2$ is plotted as black squares, and it is the only even harmonic reported. The shaded regions highlight data with a low signal-to-noise ratio. In both graphs, the vertical dashed lines mark the onset of non-linearity ($\gamma_{\rm NL}$), the yield strain ($\gamma_{y}$), and the locus of the $G''$ overshoot $(\gamma_{\rm OS})$.}
\label{fig:ak}
\end{figure*}

In contrast, the insulating CB-CMC hydrogel [Figs.~\ref{fig:Lissajous}(c)–(d)] exhibits a markedly different response. At low $\gamma_0$, the LB curves are also nearly elliptical, but the transition to non-linearity occurs more gradually and is accompanied by smoother LB curves distortion. Even at high strain amplitudes ($\gamma_0>10\%$), the $\sigma(\dot \gamma)$ LB curves remain relatively open and rounded, indicating more homogeneous yielding and increased viscous dissipation. This behavior is consistent with a soft, polymer-dominated microstructure in which CB particles are insufficiently connected to form a stress-bearing network. Eventually, for sufficiently large strain amplitudes ($\gamma_0 \gtrsim 1000\%)$, the $\sigma(\dot \gamma)$ LB curves exhibit an S-shape, yet without stress saturation, which supports a more fluid-like dissipative mechanism, compatible with a polymer-driven microstructure \cite{Hyun:2011,Rogers:2011}.

\subsubsection{Fourier Transform Rheology}
\label{sec:FTR}

In order to get more quantitative insight into the intra-cycle stress response, we turn to Fourier Transform rheology \cite{Wilhelm:2002,Hyun:2011,Ewoldt:2013b}. In particular, the stress response to the sinusoidal strain input $\gamma(t)=\gamma_0\sin(\omega t)$ can be decomposed into an elastic and a viscous component, which reads $\tau(t)=\tau_{\rm elast}(t)+\tau_{\rm visc}(t)$. This is done numerically using a Fourier decomposition of the total stress response \cite{Wilhelm:2002}, which yields:
\begin{align}
     \tau_{\rm elast}(t) =  &  \tau_0 \sum_k a_k \cos \delta_k \sin ( k \omega t ) \,,\label{eq:SigmaElas}\\
     \tau_{\rm visc}(t) = & \tau_0 \sum_k a_k \sin \delta_k \cos ( k \omega t ) \,,\label{eq:SigmaVisc}
\end{align}
where $a_k$ is the relative amplitude of the $k$-th harmonic relative to the amplitude $\tau_0$ of the fundamental (i.e., $a_1=1$) and $\delta_k$ is the phase shift of the $k$-th harmonic. The relative amplitudes $a_k$ with $k=2,\, 3,\, 5,\, 7$, and 9 are plotted against the strain amplitude $\gamma_0$ for a conductive and an insulating CB-CMC hydrogel in Figs.~\ref{fig:ak}(a) and \ref{fig:ak}(b), respectively. For both types of hydrogels, $a_2$ remains much smaller than $a_3$. This result holds for all even harmonics of higher order, but only $k=2$ is shown for clarity. The presence of even harmonics is usually attributed to artifacts such as wall slip \cite{Graham:1995,Atalik:2004}. Here, despite the use of smooth boundary conditions (see section~\ref{subsec:rheoelec} for technical details), wall slip seems to be negligible from all our experiments, except maybe at intermediate strain amplitudes (i.e., $\gamma_0 \simeq 100\%$) in the case of conductive CB-CMC hydrogels. 

The main characteristics of the non-linear stress response of CB-CMC hydrogels are encoded into the odd harmonics, and we thus limit Eqs.~\eqref{eq:SigmaElas} and~\eqref{eq:SigmaVisc} to a summation over odd indices. In the linear viscoelastic regime, odd harmonics are negligible. Here, the surprisingly high values of $a_k$ for $\gamma_0 \lesssim 0.1\%$ are due to a low signal-to-noise ratio when the stress signal is too weak (see the gray-shaded regions in Fig.~\ref{fig:ak}). 
Moreover, for both conductive and insulating CB-CMC hydrogels, the harmonics grow roughly from the onset of non-linearity, i.e., for $\gamma_0 \gtrsim \gamma_{\rm NL}$. A noticeable exception is observed for the third harmonic in the response of the insulating CB-CMC hydrogel. The latter grows for strain values $ \gamma_0 \simeq 0.1 \gamma_{\rm NL}$, well before other harmonics can be observed in the macroscopic response. 
Moreover, the two types of hydrogels show additional differences. For the conductive hydrogel, the harmonic amplitudes $a_k(\gamma_0)$ display similar strain dependence for all orders, namely a monotonic increase towards a plateau whose value decreases for increasing order $k$ [Fig.~\ref{fig:ak}(a)]. Such a simultaneous growth of non-linearities with increasing strain amplitude suggests a progressive yielding scenario, compatible with the progressive fragmentation of the network of CB particles, although potentially heterogeneous in space. In sharp contrast, the normalized amplitudes $a_k(\gamma_0)$  for insulating hydrogels all show a non-monotonic evolution that depends on the order of the harmonic [Fig.~\ref{fig:ak}(b)]. In particular, the amplitudes of the harmonics for $k=3$ and $k=5$ display an overshoot with a maximum at a strain value $\gamma_0 \simeq \gamma_{\rm OS}$ close to that of the maximum in $G''$ [see orange dashed line in Fig.~\ref{fig:ak}(b)]. This behavior hints at a polymer-dominated yielding scenario in the case of insulating CB-CMC hydrogels, in qualitative agreement with the conclusions drawn from examining the LB curves in Section~\ref {sec:LBcurves}.

\begin{figure*}[!ht]
    \centering
    \includegraphics[width=0.9\linewidth]{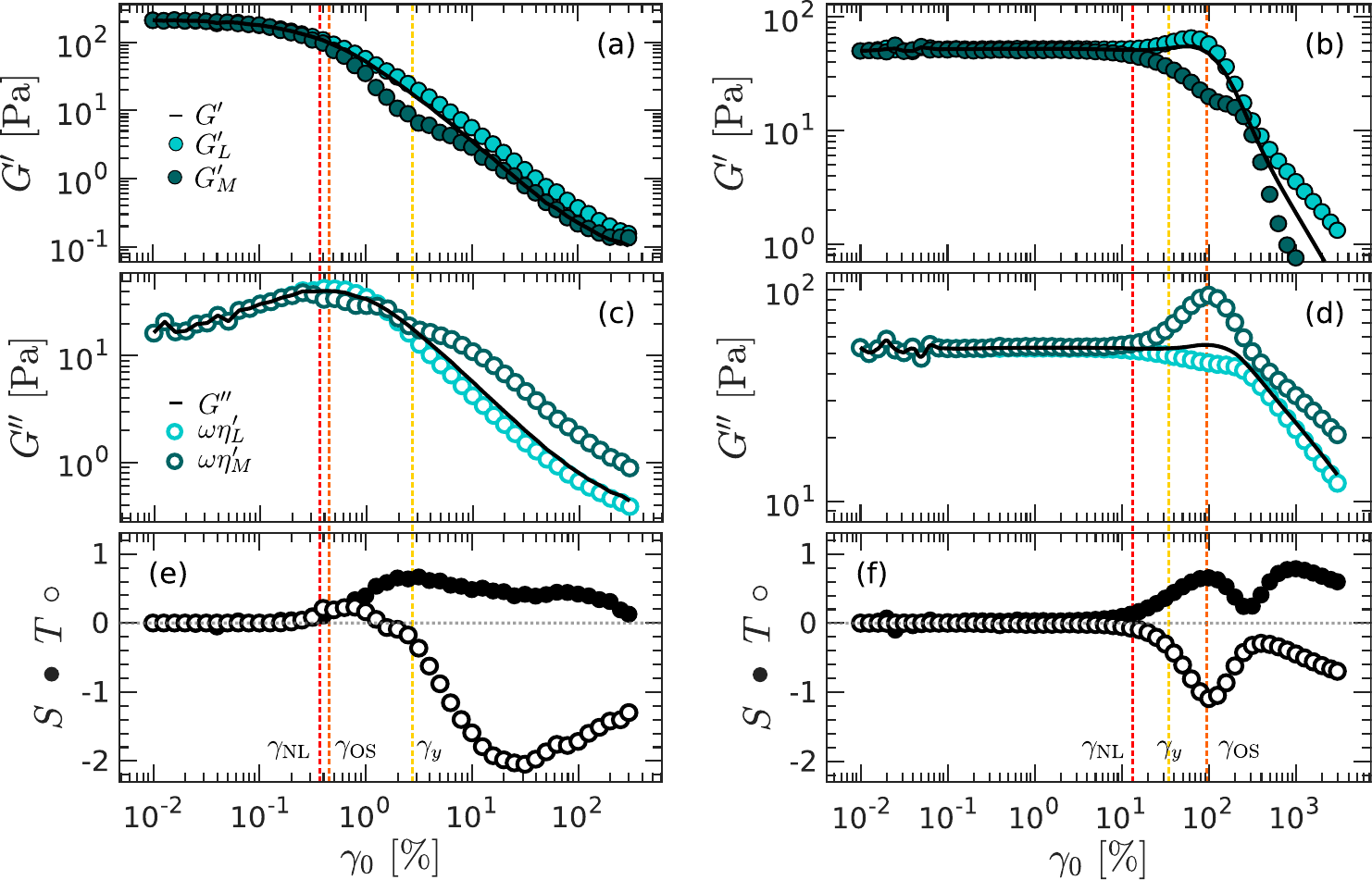}
    \caption{LAOStrain intra-cycle analysis for the two types of CB-CMC hydrogels: a conductive gel ($c_{\rm CMC} = 0.01\%$ and $x_{\rm CB} =6\%$, $r<r_c$, left column) and an insulating gel ($c_{\rm CMC} = 3\%$ and $x_{\rm CB} = 8\%$, $r>r_c$, right column). The frequency is fixed at $\omega = 2\pi~\rm rad.s^{-1}$. (a)-(b) Elastic modulus $G'$ vs.~strain amplitude $\gamma_0$ computed using only the fundamental frequency (black curve). The dark (resp. light) green disks correspond to the elastic modulus at zero strain $G'_M$ (resp. at the largest strain $G'_L$). (c)-(d) Viscous modulus $G''$ vs.~strain amplitude $\gamma_0$ computed using only the fundamental frequency (black curve). The dark (resp. light) green circles correspond to $\omega \eta'_M$ (resp. $\omega \eta'_L$). (e-f) Intra-cycle strain-stiffening parameter $S$ (filled circles) and shear-thickening parameter $T$ (open circles) vs.~$\gamma_0$. The gray horizontal dotted lines correspond to $S=T = 0$. In all graphs, the vertical dashed lines mark the onset of non-linearity ($\gamma_{\rm NL}$), the yield strain ($\gamma_{y}$), and the strain at which $G''$ reaches a maximum ($\gamma_{\rm OS}$). 
    }
\label{fig:strain_sweep_intracycle_analysis}
\end{figure*}

\subsubsection{Chebyshev decomposition}

A complementary approach to Fourier Transform rheology, pioneered in refs.~\cite{Cho:2005,Ewoldt:2008} considers the decomposition of the non-linear stress response into Chebyshev polynomials. The elastic and viscous stresses, $\tau_{\rm elast} (\gamma)$ and $\tau_{\rm visc}(\dot{\gamma})$, are expressed as follows:
\begin{align}
    \tau_{\rm elast}(\gamma) &= \gamma_0 \sum_{k, \ \text{odd}} e_k T_k \left( \dfrac{\gamma}{ \gamma_0} \right) \,, \\
    \tau_{\rm visc}(\dot \gamma) &= \omega \gamma_0 \sum_{k, \ \text{odd}} v_k T_k \left( \dfrac{\dot{\gamma}}{ \omega \gamma}_0 \right) \,,
\end{align}
with $T_k$ the Chebyshev polynomials of order $k$, and $e_k$ and $v_k$ the coefficients of order $k$ for the elastic and viscous decomposition, respectively. In practice, these coefficients are computed numerically through a scalar product of the data with each $T_k$ for the first 3 terms, i.e., $k=1,\, 3,\, \rm and\,\, 5$. 
This approach allows one to compute two intra-cycle elastic moduli  \cite{Kamkar:2022} :
\begin{align}
    G'_L &= \left. \dfrac{\tau_{\rm elast} }{\gamma} \right| _{\gamma = \gamma_0}  = \sum_{k, \ \text{odd}} e_k \label{eq:mat_meth_GL} \,, \\
    G'_M &= \left. \dfrac{\mathrm{d} \ \tau_{\rm elast} }{\mathrm{d} \gamma} \right| _{\gamma = 0} = \sum_{k, \ \text{odd}} k (-1)^{(k-1)/2}   e_k \,.   \label{eq:mat_meth_GM}
\end{align}
$G'_L$ and $G'_M$ are respectively the modulus at the largest strain within a cycle, and the modulus at zero strain. These moduli are reported as a function of the strain amplitude $\gamma_0$ in Figs.~\ref{fig:strain_sweep_intracycle_analysis}(a) and \ref{fig:strain_sweep_intracycle_analysis}(b) for conductive and insulating hydrogels, respectively. We also introduce the minimum-rate dynamic viscosity $\eta'_M$ and large-rate dynamic viscosity $\eta'_L$, which are defined as follows \cite{Kamkar:2022}:
\begin{align}
    \eta'_L &= \left. \dfrac{\tau_{\rm visc}}{\dot{\gamma}} \right| _{\dot{\gamma} = \omega \gamma_0} = \sum_{k, \ \text{odd}} v_k \label{eq:mat_meth_etaL} \\
    \eta'_M &= \left. \dfrac{\mathrm{d} \ \tau_{\rm visc} }{\mathrm{d} \dot{\gamma} }\right| _{ \dot{\gamma} = 0 }  = \sum_{k, \ \text{odd}} k (-1)^{(k-1)/2}  v_k \label{eq:mat_meth_etaM}
\end{align}
and reported in Figs.~\ref{fig:strain_sweep_intracycle_analysis}(c) and \ref{fig:strain_sweep_intracycle_analysis}(d) vs $\gamma_0$ for conductive and insulating hydrogels, respectively. Finally, these quantities allow introducing two dimensionless parameters to quantify the relative amount of intra-cycle strain-stiffening $S=(G'_L-G'_M)/G'_L$, and shear-thickening $T=(\eta'_L-\eta'_M)/\eta'_L$ \cite{Kamkar:2022}. The parameters $S$ and $T$ are reported as a function of the strain amplitude $\gamma_0$ in Fig.~\ref{fig:strain_sweep_intracycle_analysis}(e) and \ref{fig:strain_sweep_intracycle_analysis}(f) for conductive and insulating hydrogels, respectively. In the linear deformation regime, $G'_L=G'_M=G'$, and $\eta'_L=\eta'_M=G''/\omega$, while $S=T=0$. 
In the non-linear regime, the Chebyshev approach provides more physical insight than Fourier Transform rheology by focusing on only a few intra-cycle parameters that convey the impact of the full harmonics series within the stress response \cite{Ewoldt:2008,Ewoldt:2013}.

Let us first examine the case of conductive CB-CMC hydrogels, which corresponds to the left column in  Fig.~\ref{fig:strain_sweep_intracycle_analysis}. 
We observe that, for $\gamma_0 \gtrsim \gamma_{\rm NL}$, i.e., throughout the entire non-linear regime, the elastic modulus $G'_L$ defined at the moment of maximum deformation in the cycle remains larger than $G'_M$ defined at $\gamma=0$ in the cycle. This behavior is well-summarized by the strain dependence of the strain-stiffening index $S$, which is zero in the linear regime, and becomes positive for $\gamma_0 \gtrsim \gamma_{\rm NL}$, indicating an intra-cycle strain-stiffening behavior of the conductive hydrogel [Fig.~\ref{fig:strain_sweep_intracycle_analysis}(e)]. This parameter keeps growing for increasing strain amplitude up to a maximum $S\simeq 0.6$ reached at $\gamma_0 \simeq \gamma_y$. Beyond this point, $S(\gamma_0)$ slowly decreases while remaining positive, showing that the conductive hydrogel displays intra-cycle strain stiffening deep into the non-linear regime, well beyond the yield point.  
Concomitantly, the large-rate dynamic viscosity $\eta'_L$ defined at the maximum shear rate $\dot \gamma = \omega \gamma_0$ coincides well with $G''/\omega$ over the entire range of strain amplitude explored [Fig.~\ref{fig:strain_sweep_intracycle_analysis}(c)]. However, while $\eta'_L > \eta'_M$ up to $\gamma_0 \simeq \gamma_y$, we observe that $\eta'_L < \eta'_M$ beyond the yield point, i.e., that the intra-cycle viscosity defined at zero shear becomes the largest. In other words, the conductive hydrogel displays a mildly intra-cycle shear-thickening response ($T>0$) for $\gamma_{\rm NL} < \gamma < \gamma_y$, followed by a much more pronounced intra-cycle shear-thinning response ($T<0$) for $\gamma_0 \gtrsim \gamma_y$ [Fig.~\ref{fig:strain_sweep_intracycle_analysis}(e)]. Additionally, the shear-thickening index reaches a well-defined minimum $T\simeq -2$ at $\gamma_0 \simeq 30\% \gg \gamma_y$ beyond which it exhibits a slow increase. This change in the shear-thickening index is not obviously reflected in the strain-stiffening one. 
In summary, these results suggest the following microscopic scenario underpinning the LAOStrain response of conductive CB-CMC hydrogels: as the strain amplitude increases, the percolated network of CB particles exhibits a progressively higher level of resistance to deformation up to the yield point, where the CB network likely breaks into clusters. These clusters still display some resistance to shear, which explains why the strain-stiffening index remains positive well beyond the yield point. However, they do not form a percolated network and may align with the flow, resulting in a pronounced intra-cycle shear-thinning response. However, some dramatic microstructural change must occur at $\gamma_0 \simeq 30\%$ beyond which $T$ increases, which will be interpreted in light of the time-resolved rheo-electrical measurements reported in Section~\ref{sec:RheoElecConduc}. 

We now turn to the case of insulating CB-CMC hydrogels, which is illustrated in the right column in Fig.~\ref{fig:strain_sweep_intracycle_analysis}. In the linear regime ($\gamma_0 < \gamma_{\rm NL}$), $G' = G'_M=G'_M$ and $S=T=0$, as expected. In the non-linear regime, i.e., for $\gamma_0 \geqslant \gamma_{\rm NL}$, $G'$ roughly coincides with $G'_L$, the modulus defined at the moment of maximum deformation in the cycle, up to $\gamma_0 \simeq 300\%$, beyond which $G'_M< G'<G'_L$ [Fig.~\ref{fig:strain_sweep_intracycle_analysis}(b)]. Moreover, $G'_M<G'_L$ for the entire range of non-linear strain explored, which means that the insulating CB-CMC hydrogel mainly exhibits intra-cycle strain stiffening, as evidenced by $S>0$, despite a complex, non-monotonic evolution [Fig.~\ref{fig:strain_sweep_intracycle_analysis}(f)]. Interestingly, the strain-stiffening index $S$ keeps increasing beyond the yield point, showing a clear maximum together with $G''$ at $\gamma = \gamma_{\rm OS}$ [in stark contrast with the conductive sample where the maximum of $S$ is linked to $\gamma_y$, see Fig.~\ref{fig:strain_sweep_intracycle_analysis}(e)]. Beyond that point, $S$ shows a local minimum at $\gamma_0 \simeq 300\%$ followed by a local maximum at $\gamma_0 \simeq 1000\%$ beyond which $S(\gamma_0)$ shows a weakly decaying trend.
Concomitantly, for $\gamma_0 \geqslant \gamma_{\rm NL}$, the large rate dynamic viscosity $\eta'_L$ coincides reasonably well with $G''/\omega$, except in the vicinity of the overshoot in $G''$, where $\eta'_M>G''/\omega > \eta'_L$. Throughout the non-linear regime, $\eta'_M>\eta'_L$, which means that the insulating CB-CMC hydrogels only display intra-cycle shear-thinning, as captured by the negative values of the shear-thickening index $T<0$ [Fig.~\ref{fig:strain_sweep_intracycle_analysis}(f)]. In addition, $T(\gamma_0)$ is anti-correlated with $S(\gamma_0)$, reaching a minimum $T \simeq -1.2$ at $\gamma_0=\gamma_{\rm OS}$, while the strain-stiffening index $S$ is maximum; then $T$ increases up to $T \simeq -0.2$ reached at $\gamma_0 \simeq 300\%$ where $S$ is maximum, beyond which $T$ shows a monotonic decay. 
This phenomenology agrees with a polymer-dominated microstructure, in which the CMC polymers become progressively aligned along the flow direction for increasing strain amplitude ($T<0$) while resisting deformation ($S>0$) due to CB particles playing the role of physical crosslinkers, and due to polymer entanglement and/or polymer chain finite extensibility \cite{nebouy2021flow,Kamkar:2022,Legrand:2025}. Finally, the local maximum in $T$ and minimum in $S$ observed for larger strain amplitudes at $\gamma_0 \simeq 300\%$ remains unclear at this stage, and is further interpreted by time-resolved rheo-USAXS measurements reported in Section~\ref{sec:RheoTRUSAXS}.

These results confirm the fundamentally different scenarios underlying the non-linear response of conductive and insulating CB-CMC hydrogels to large-amplitude oscillations. In conductive CB-CMC hydrogels, the percolated CB network is progressively disrupted while resisting external deformation up to the yield point, beyond which the sample behaves as a shear-thinning suspension of isolated clusters of CB particles. In contrast, the response of insulating CB-CMC hydrogels is driven by the CMC polymers, which align along the flow direction from the onset of the non-linear regime, while CB particles might play a role at larger strain amplitudes. To further explore the non-linear response of these hydrogels, we now turn to time-resolved electric and USAXS measurements coupled with LAOStrain.

\subsection{Insights from time-resolved rheo-electric experiments on conductive CB-CMC hydrogels}
\label{sec:RheoElecConduc}

To probe the evolution of the hydrogel microstructure under shear, we monitor the electrical properties of conductive CB-CMC hydrogels during a LAOStrain experiment using the rheo-electric setup described in section~\ref{subsec:rheoelec}. Here, we apply a constant voltage $U_0 = 100~\rm mV$ and record the current $I$ to compute the effective instantaneous electrical DC conductivity $\sigma_{\rm DC}(t) = k I(t) / U_0$, with $k$ the cell constant. This value is a direct image of the connectivity of the CB percolated network since the electrical response time (around 10~ms for the colloidal gel) is much shorter than the timescale of the strain sweep, set by the frequency of mechanical oscillations of about $1~\rm s$ \cite{legrand:tel-04768650}.

\begin{figure*}[!th]
  \begin{minipage}{0.48\textwidth}
    \centering
    \includegraphics[width=0.95\linewidth]{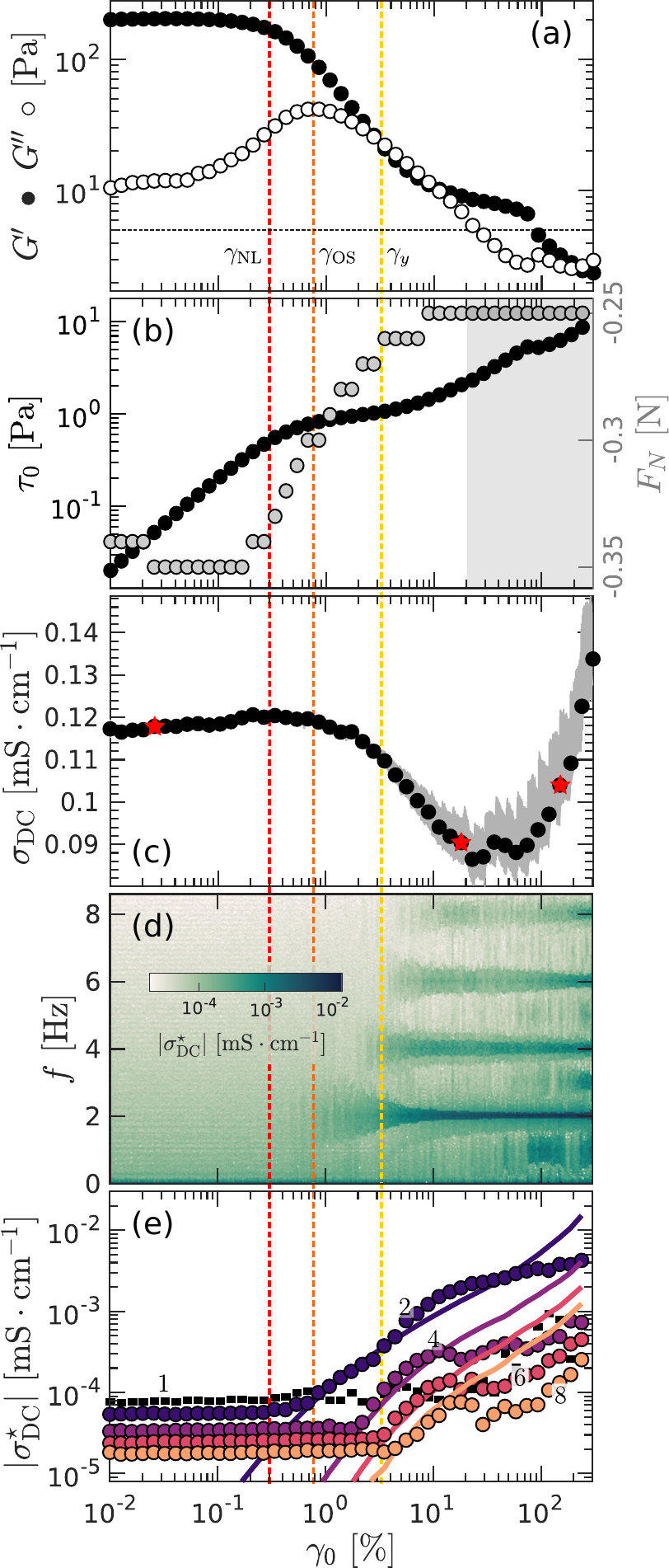}
  \end{minipage}\hfill
  \begin{minipage}{0.48\textwidth}
    \caption{LAOStrain experiments on a conductive CB-CMC hydrogel ($c_{\rm CMC} = 0.01\%$ and $x_{\rm CB} = 6 \%$, $r<r_c$) performed with the rheo-electric setup at $\omega = 2\pi ~\rm rad.s^{-1}$. Data are averaged over $2$ cycles per step. Vertical dashed lines indicate $\gamma_{\rm NL}$ (red), $\gamma_{\rm OS}$ (orange), and $\gamma_{y}$ (yellow) as defined in Fig.~\ref{fig:strain_sweep_CB8}. 
(a) Elastic modulus $G'$ (full symbols) and viscous modulus $G''$ (empty symbols) vs.~$\gamma_0$. The horizontal dashed line shows the limit below which mechanical measurements are affected by the electrical wire (see Appendix~\ref{appendix:rheoelec}). (b) Stress amplitude $\tau_0$ and normal force $F_N$ vs.~$\gamma_0$. (c) Electrical conductivity $\sigma_{\rm DC}$ vs.~$\gamma_0$: raw data (sampled at 100~Hz) are shown in gray, while the average over each mechanical cycle --corresponding to the zero-frequency component in (d)-- are shown as black disks. Red stars mark the segment shown in Fig.~\ref{fig:zoom_conducti} in Appendix~\ref{Appendix:ElecVariation}. (d) Strain-frequency diagram of $|\sigma_{\rm DC}^\star|$, the modulus of the Fourier transform $\sigma_{\rm DC}^\star$ of the electrical conductivity $\sigma_{\rm DC}(t)$ computed over two cycles for each value of $\gamma_0$.
Lighter colors indicate larger amplitudes (see color bar). (e) Harmonic amplitudes $ | \sigma_{\rm DC, p}^\star | =|\sigma_{\rm DC}^\star(f= p \omega/2\pi)|$ vs.~$\gamma_0$ with $p= 2$, 4, 6, and 8 as indicated on the graph. These harmonics correspond to the darkest horizontal lines in (d). The $p=1$ harmonic (odd harmonic of distinct physical meaning) is plotted as black squares. The curves correspond to the best fits of the data with Eq.~\eqref{eq:harmonics}.}
    \label{fig:rheo_elec_coll}
  \end{minipage}
\end{figure*}

The results of a representative rheo-electric experiment are shown in Fig.~\ref{fig:rheo_elec_coll} for the same conductive CB-CMC hydrogel as that already studied above in Figs.~\ref{fig:strain_sweep_CB8}(a,c), \ref{fig:Lissajous}(a,b), \ref{fig:ak}(a), and \ref{fig:strain_sweep_intracycle_analysis}(a,c,e). First, the rheological response measured with the rheo-electric device in a parallel-plate geometry displays qualitatively the same key features as the response measured in a cone-and-plate geometry [compare Fig.~\ref{fig:rheo_elec_coll}(a) with Fig.~\ref{fig:strain_sweep_CB8}(a) and Fig.~\ref{fig:strain_sweep_intracycle_analysis}(a)], confirming the robustness of the rheological response of the conductive CB-CMC hydrogel. The main difference is that the rheo-electric device does not allow us to measure viscoelastic modulus lower than $5~\rm Pa$, as discussed in Appendix~\ref{appendix:rheoelec} [see horizontal dashed line in Fig.~\ref{fig:rheo_elec_coll}(a)]. We also leverage the parallel-plate geometry to report the normal force $F_N$.

The key result lies in the simultaneous measurement of the DC electrical conductivity, $\sigma_{\rm DC }$, reported in Fig.~\ref{fig:rheo_elec_coll}(c). Although the different strain amplitudes are applied stepwise, the electric signal $\sigma_{\rm DC }(t)$ is plotted continuously as a function of $\gamma_0(t)$. The conductivity remains approximately constant from the linear deformation regime up to the strain amplitude $\gamma_{\rm OS}$ corresponding to the peak in $G''$. For strain amplitudes $\gamma_0 \gtrsim \gamma_{\rm OS}$, the conductivity progressively decreases, hence confirming the disruption of the percolated CB network under large amplitude oscillatory shear. Moreover, a minimum in conductivity is observed at $\gamma_0 \simeq 30\%$, beyond which $\sigma_{\rm DC}$ increases with increasing strain amplitude, eventually overshooting the original conductivity value of the conductive CB-CMC hydrogel at rest. The conductivity rise occurs while the sample exhibits a liquid-like response, with $G' <G''$. Concomitantly, the measured normal force only shows a moderate increase between $\gamma_{\rm NL}$ and $\gamma_y$ [see Fig.~\ref{fig:rheo_elec_coll}(b)], in contrast with denser colloidal systems \cite{Fournier:2024}. These observations suggest that the clusters of CB particles inherited from the broken CB network might form a shear-induced dynamic network with enhanced electrical conductivity due to shear. Similar increases in conductivity under shear have been reported in pure CB gels in mineral oil during creep experiments \cite{Helal:2016}, as well as in molecular dynamics simulations of colloidal gels under shear start-up \cite{Colombo:2014}. Importantly, because the electrical percolation threshold is known to be much lower than the mechanical percolation threshold in CB gels \cite{Richards:2017}, the observed rise in electrical conductivity does not necessarily coincide with major changes in the rheological properties.
Finally, we verified the transient nature of the shear-induced network by abruptly stopping the shear, which results in an abrupt drop of the conductivity to a value that is the one measured at rest before the start of the strain sweep (see Fig.~\ref{fig:relax} in Appendix~\ref{Appendix:Relaxation}).   

\begin{figure*}[!ht]
    \centering
    \includegraphics[width=0.85\linewidth]{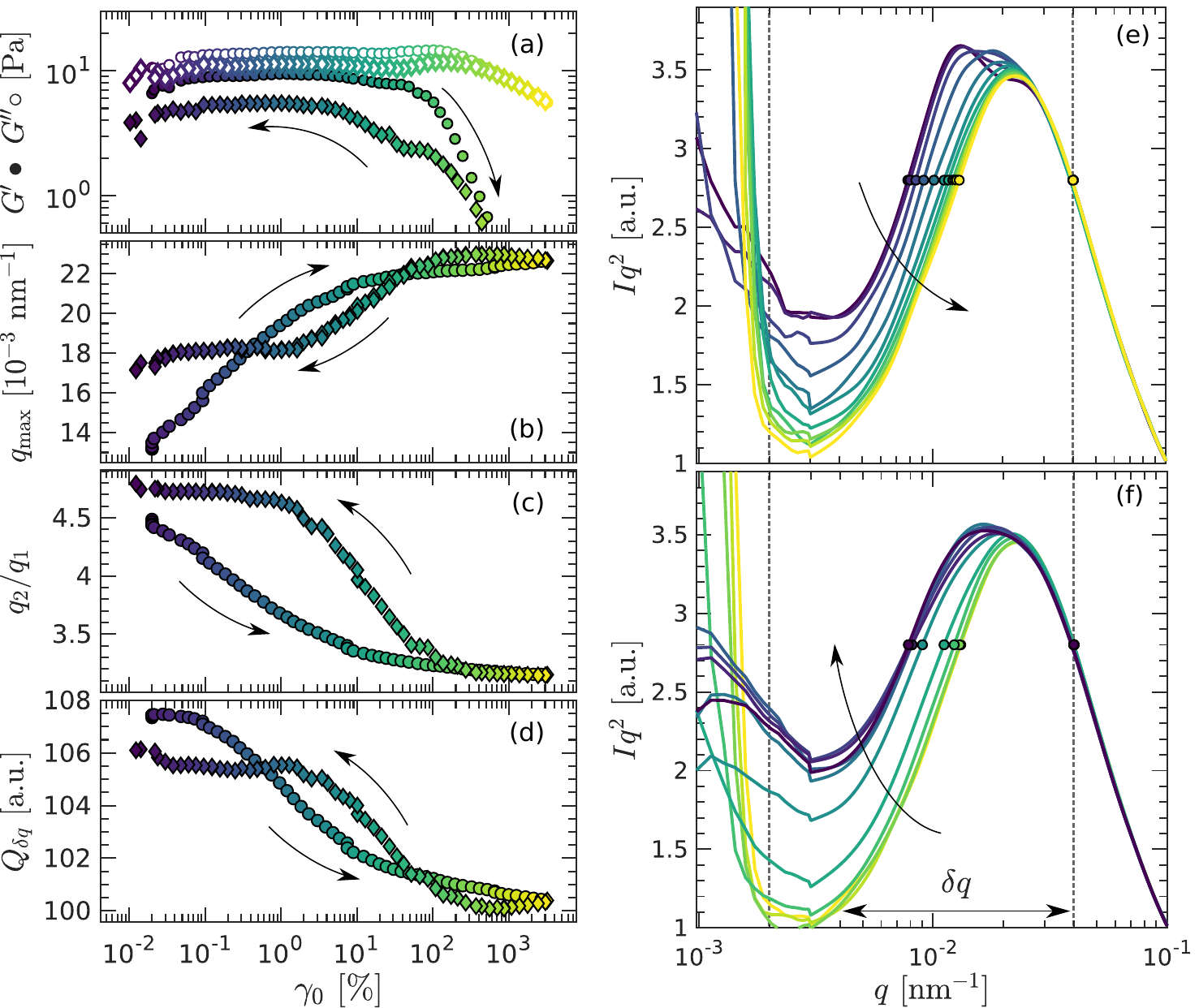}
    \caption{LAOStrain experiments performed using rheo-TRUSAXS on an insulating CB-CMC hydrogel ($c_{\rm CMC} = 2\%$ and $x_{\rm CB} = 8\%$, $r>r_c$). The frequency is set to $\omega = 2\pi~\rm rad.s^{-1}$, and the strain amplitude $\gamma_0$ is progressively increased from $0.01~\%$ to $3 000~\%$ (circles), then decreased over the same range (diamonds), as indicated by the black arrows. (a) Elastic modulus $G'$ (filled symbols) and viscous modulus $G''$ (empty symbols) moduli as a function of $\gamma_0$. (b) Scattering wave vector $q_{\rm max}$ corresponding to the local maximum in the Kratky plot $Iq^2$ vs.~$q$ and plotted against $\gamma_0$. (c) Width of the peak $q_2 / q_1$ as a function of $\gamma_0$, where $q_2 = 4.10^{-2} \rm ~mm^{-1}$ and $q_1<q_2$ is the largest scattering wave vector below $q_{\rm max}$ that corresponds to $I(q_2)q_2^2$ in the Kratky plot, and indicated as colored dots. (d) Area $Q_{\delta q}$ enclosed by the Kratky plot between the two vertical dashed lines shown in (e) and (f) and plotted against $\gamma_0$. (e,f) Kratky plots $Iq^2$ vs.~$q$ during (e) the increasing strain ramp and (f) decreasing strain ramp. Colors indicate the strain amplitude $\gamma_0$, with lighter colors representing larger values of $\gamma_0$, as shown by the symbol colors in (a--d).}
    \label{fig:rheo_saxs_polym}
\end{figure*}

A closer examination of the conductivity signal as a function of the strain amplitude reveals an increasing level of fluctuations beyond the yield point, as visible in the raw data shown in gray in Fig.~\ref{fig:rheo_elec_coll}(c) and in the zoomed traces of this signal for strain values highlighted here by red stars, and reported in Fig.~\ref{fig:zoom_conducti} in Appendix~\ref{Appendix:ElecVariation}. The corresponding strain-frequency diagram from a Fourier analysis, presented in Fig.~\ref{fig:rheo_elec_coll}(d), reveals the progressive emergence of even harmonics with increasing strain amplitude. Specifically, the second harmonic appears starting from $\gamma_0=\gamma_{\rm OS}$, followed by the emergence of higher even harmonics, $p=4,\, 6,$ and $8$ at $\gamma_0=\gamma_y$, marking the yield point [see Fig.~\ref{fig:rheo_elec_coll}(e)].
Since the electrical conductivity is measured under a constant voltage (i.e., with zero frequency), the harmonic content in the electrical signal must originate from the imposed mechanical oscillations. Since electrical conduction is a priori insensitive to the direction of shear, it is natural to express $\sigma_{\rm DC}$ as a function of the instantaneous \textit{mechanical} power $\mathcal{P} = \dot \gamma \tau $ injected into the system.
Expanding the conductivity in a Taylor series for small $\mathcal{P}$, i.e., for low strain amplitudes, yields:
$\sigma_{\rm DC} = c_{0} + c_{1} \mathcal{P} + \mathcal{O}(\mathcal{P}^{2})$
where $c_{0}$ corresponds to the mean conductivity [see black disks in Fig.~\ref{fig:rheo_elec_coll}(c)]. For a sinusoidal strain input $\gamma(t)=\gamma_0\sin(\omega t)$, and a stress waveform output written using the notations introduced in Section~\ref{sec:FTR}:
\begin{align}
    \tau(t) &= \tau_0 \sum_k  a_k \sin \left(  k \omega t + \delta_k \right) \,,\label{eq:stress}
\end{align}
one can compute the linear term in $\mathcal{P}$, which can be expressed after some algebra as:
\begin{align}
    \mathcal{P}(t)=  \frac{1}{2} \omega \gamma_0 \tau_0 \sum_n b_{n} \sin \left(  n \omega t - \phi_{n} \right)\,,
\end{align}
where the amplitudes $b_{n}$ and phases $\phi_{n}$ of the electrical signal are functions of $a_j$ and $\delta_j$, with $j=n-1$, $n$, and $n+1$ (see Appendix~\ref{Appendix:calc} for full expressions). Experimentally, we observe that $\sigma_{\rm DC}$ exhibits only even harmonics of the mechanical excitation at the frequency $\omega$, which is consistent with the fact that $a_{n\pm1} \simeq 0$ for odd $n$, leading to $b_n\simeq 0$. This supports the interpretation that the even harmonics in the electrical signal arise from the convolution of neighboring odd harmonics in the stress. As detailed in Appendix~\ref{Appendix:calc}, up to first order in $\mathcal{P}$, i.e., under the assumption that $\sigma_{\rm DC} = c_{0} + c_{1} \mathcal{P}$, the amplitudes of the harmonics of the electrical signal are linked to the non-linear rheological response as follows:
\begin{align}
    | \sigma_{\rm DC, n}^\star | = \frac{1}{2} c_1 |G^\star| \omega \gamma_0^2 b_n \,,\label{eq:harmonics}
\end{align}
where $c_1$ is a constant that neither depends on $\gamma_0$ nor on $n$, $b_n \simeq a_{n-1}$ with exact expression provided in Eq.~\eqref{eq.bn} in Appendix~\ref{Appendix:calc}, and starred quantities indicate Fourier transforms. The best fit of the data using Eq.~\eqref{eq:harmonics} is shown in Fig.~\ref{fig:rheo_elec_coll}(e) and displays an excellent agreement with the data, at least up to $\gamma_0 \simeq 20\%$, beyond which the underlying assumption of small $\mathcal{P}$ most probably fails. 

In conclusion, the coupling between mechanical excitation and electrical response in CB-CMC hydrogels reveals rich nonlinear dynamics, including the formation of a transient conductive network under shear. To investigate the changes in the sample microstructure, we performed time-resolved rheo-TRUSAXS experiments. The results, summarized in Fig.~\ref{fig:rheo_saxs_coll} and reported in Appendix~\ref{Appendix:TRUSAXSconductive}, show that the scattering intensity does not display any significant change during the LAOStrain experiment. These results demonstrate that the yielding transition occurs at length scales that lie outside the window of wave vectors accessible to USAXS, hence above a few microns.

\subsection{Insights from rheo-TRUSAXS and time-resolved rheo-electric experiments on insulating CB-CMC hydrogels}
\label{sec:RheoTRUSAXS}

\begin{figure*}[!th]
  \begin{minipage}{0.48\textwidth}
    \centering
    \includegraphics[width=0.95\linewidth]{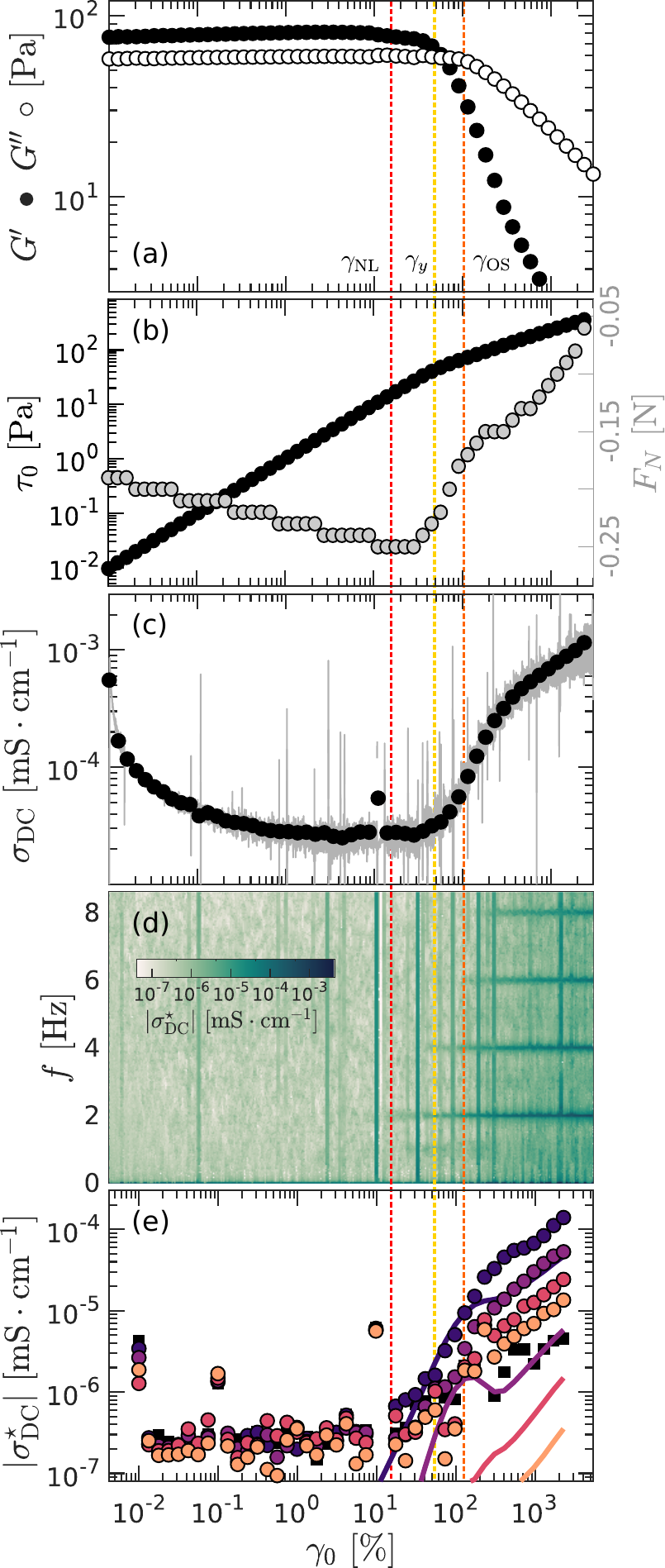}
  \end{minipage}\hfill
  \begin{minipage}{0.48\textwidth}
    \caption{LAOStrain experiment on an insulating CB-CMC hydrogel ($c_{\rm CMC} = 3\%$ and $x_{\rm CB} = 8\%$) performed with the rheo-electric setup at $\omega = 2\pi ~\rm rad.s^{-1}$. Data are averaged over $2$ cycles per step. Vertical dashed lines indicate $\gamma_{\rm NL}$ (red), $\gamma_{\rm OS}$ (orange), and $\gamma_y$ (yellow) as defined in Fig.~\ref{fig:strain_sweep_CB8}. (a) Elastic modulus $G'$ (full symbols) and viscous modulus $G''$ (empty symbols) vs.~$\gamma_0$. (b) Stress amplitude $\tau_0$ and normal force $F_N$ vs $\gamma_0$. (c) Electrical conductivity $\sigma_{\rm DC}$ vs.~$\gamma_0$: raw data (sampled at $100~\rm Hz$) are shown in gray, while the average over each oscillation cycle --corresponding to the zero-frequency component in (d)-- are shown as black disks. (d) Strain-frequency diagram of $|\sigma_{\rm DC}^\star|$, the modulus of the Fourier transform $\sigma_{\rm DC}^\star$ of the electrical conductivity $\sigma_{\rm DC}(t)$ computed over two cycles for each value of $\gamma_0$. Darker colors indicate larger amplitudes (see color bar). (d) Harmonic amplitudes $ | \sigma_{\rm DC, p}^\star | =|\sigma_{\rm DC}^\star(f= p \omega/2\pi)|$ vs.~$\gamma_0$ with $p=1,\, 2,\, 4,\, 6,\, \rm and\, 8$, as indicated on the graph. The $p=1$ harmonic (odd harmonic of distinct physical meaning) is plotted in black squares. The curves correspond to the best fits of the data with Eq.~\eqref{eq:harmonics}.}
    \label{fig:rheo_elec_polym}
  \end{minipage}
\end{figure*}

In this section, we explore the microstructural evolution of the insulating CB-CMC hydrogels already studied above in Figs.~\ref{fig:strain_sweep_CB8}(b,d), \ref{fig:Lissajous}(c,d), \ref{fig:ak}(b), and \ref{fig:strain_sweep_intracycle_analysis}(b,d,f) during the yielding transition induced by LAOStrain. We first report on rheo-TRUSAXS and the result of a strain sweep of increasing amplitude immediately followed by a ramp of decreasing amplitude over the same range of strains is reported in Fig.~\ref{fig:rheo_saxs_polym}. The strain sweep is performed at a fixed frequency $\omega = 2\pi~\rm rad.s^{-1}$ on a weakly aged sample, i.e., 3~min after rejuvenation imposed by the preshear (see Section~\ref{subsec:rheometry} for the reference rheology protocol). This choice was made due to time constraints when performing the experiments at ESRF. Although $G'<G''$ at this frequency [see Fig.~\ref{fig:rheo_saxs_polym}(a)], the sample is indeed a soft solid following such a short waiting time since $G'>G''$ in the low-frequency limit \cite{Legrand:2023}.

Despite this specificity, the sample gets fluidized for increasing strain amplitude since $G''$ shows an overshoot with a peak at $\gamma_0 = \gamma_{\rm OS} \simeq 100\%$, while $G'$ exhibits a sharp decay for $\gamma_0 \gtrsim \gamma_{\rm OS}$ in agreement with our previous observations on older samples tested in other shear geometries [see Figs.~\ref{fig:strain_sweep_CB8}(b) and \ref{fig:strain_sweep_intracycle_analysis}(b,d)]. 
The evolution of the scattering intensity is reported in Fig.~\ref{fig:rheo_saxs_polym}(e,f) as a Kratky plot, i.e., $Iq^2$ vs.~$q$, upon increasing and decreasing the shear amplitude, respectively. In that representation, the intensity shows a third-order polynomial shape similar to that obtained in static measurements performed on capillaries and reported in Fig.~\ref{fig:figure1}. We remind the reader that the local maximum located at $q=q_{\rm max}$ originates from both the diameter of CB particles and the most probable distance that separates their centers of mass (see discussion in Section~\ref{sec:Results0}). For $q \gtrsim 2.10^{-2}~\rm nm^{-1}$, $Iq^2$ is insensitive to the strain amplitude, since it corresponds to the intensity scattered by the surface of CB particles. In contrast, for $q \lesssim 2.10^{-2}~\rm nm^{-1}$, $Iq^2$ shows a continuous evolution for increasing strain amplitude. In particular, the position of the maximum $q_{\rm max}$ shifts towards larger values from $1.4\cdot 10^{-2}~\rm nm^{-1}$ to $2.2\cdot 10^{-2}~\rm nm^{-1}$ [Fig.~\ref{fig:rheo_saxs_polym}(b)], which is the signature of the fragmentation of CB clusters under shear. Simultaneously, the width of the peak decreases for increasing strain amplitude, as illustrated by the ratio $q_2/q_1$ reported in Fig.~\ref{fig:rheo_saxs_polym}(c). This indicates a narrowing of the size distribution of the CB clusters --the shear playing the role of a low-pass filter. Eventually, both $q_{\rm max}$ and the ratio $q_2 /q_1$ become constant, for $\gamma_0 \simeq 100\% \simeq \gamma_{\rm OS}$. Interestingly, the scattering data do not display any obvious signature of the yield point, previously identified at $\gamma_y \simeq 70\%$ for this sample composition [see Fig.~\ref{fig:strain_sweep_CB8}(b) and \ref{fig:strain_sweep_CB8}(d)]. Indeed, one should recall that the scattering intensity originates from the CB particles, whereas the rheological properties are mainly driven by the CMC polymer matrix.   

Besides the presence of a local maximum, we also observe a strong increase in $Iq^2$ at $q \leq 2.10^{-3}~\rm nm^{-1}$ when increasing the strain amplitude [see Fig.~\ref{fig:rheo_saxs_polym}(e)]. This result suggests the formation of strain-induced larger-scale structures or even a transient percolated network made of fine CB clusters. This scenario is further supported by examining the invariant $Q_{\rm SAXS}$, which represents the area under the Kratky plot over the entire spectrum and must be constant as the system composition does not change during the experiment \cite{Lindner:1991}. In Fig.~\ref{fig:rheo_saxs_polym}(d), we compute a fraction of this area $Q_{\delta q}$ by integrating $Iq^2$ over the range of scattering wave vectors comprised between the vertical dashed lines in Fig.~\ref{fig:rheo_saxs_polym}(e). Because the Kratky plot is strain-independent for $q$ values larger than in the region considered for $Q_{\delta q}$ (i.e., $q\gtrsim 4.10^{-2}~\rm nm^{-1}$), the decrease of $Q_{\delta q}$ as $\gamma_0$ increases must be compensated by an increase in the scattering intensity at lower $q$ values, outside the $q$-range considered for $Q_{\delta q}$ (i.e., $q\lesssim 2.10^{-3}~\rm nm^{-1}$). From the data available at the lowest $q$ values, this appears to be the case. This important result rules out any artifact from scattering at low $q$ due to, e.g., bubbles or crazes, and confirms a scenario involving the strain-induced formation of larger structures. 

The microstructural scenario inferred from rheo-TRUSAXS is further confirmed by the time-resolved rheo-electric experiments summarized in Fig.~\ref{fig:rheo_elec_polym}. The evolution of the viscoelastic moduli is displayed in Fig.~\ref{fig:rheo_elec_polym}(a) and consistent with the phenomenology reported for other compositions of insulating CB-CMC hydrogels in Fig.~\ref{fig:strain_sweep_CB8}(b). Interestingly, beyond the yield point, i.e., for $\gamma_0 \gtrsim \gamma_y$, the normal force displays a relative increase of $0.2~\rm N$ that extends deep in the non-linear regime, in contrast with conductive hydrogels [see Fig.~\ref{fig:rheo_elec_coll}]. Concomitantly, there is a significant increase in the DC conductivity by an order of magnitude, which confirms that shear tends to bring the CB particles and clusters closer together, favoring the formation of a more connected CB network. Nonetheless, the absolute value of the conductivity remains much smaller than that measured in conductive CB-CMC hydrogels, at best $\sigma_{\rm DC} \simeq 1~\rm\mu S.cm^{-1}$ compared to $0.1~\rm mS.cm^{-1}$ [see Fig.~\ref{fig:rheo_elec_coll}(c)], which rules out the existence of a space spanning network. Note that the electrical conductivity also exhibits fluctuation, like in the case of the conductive CB-CMC hydrogels, and that the strain-frequency diagram also shows even harmonics [see Fig.~\ref{fig:rheo_elec_polym}(d)] due to the coupling with mechanical oscillations. Last but not least, the increase in conductivity is only transient, for the conductivity drops to its initial value upon flow cessation [see Fig.~\ref{fig:relax}(b) in Appendix~\ref{Appendix:Relaxation}]. 

Coming back to the rheo-TRUSAXS experiments reported in  Fig.~\ref{fig:rheo_saxs_polym}, we now examine the response of the insulating CB-CMC hydrogel to a strain sweep of decreasing amplitude immediately following the above-discussed increasing ramp [see diamonds in Fig.~\ref{fig:rheo_saxs_polym}(a)-(d) and Fig.~\ref{fig:rheo_saxs_polym}(f)]. Starting from large strain values, $\gamma_0 \simeq 1000\%$, the sample is liquid-like ($G'' \gg G'$). As the strain amplitude decreases, $G'$ increases for $\gamma_0 \lesssim 500\%$, showing that the gel progressively reforms [see Fig.~\ref{fig:rheo_saxs_polym}(a)], while the scattering data $Iq^2$ show the opposite trend to that observed during the increasing ramp [see Fig.~\ref{fig:rheo_saxs_polym}(f)]. More specifically, $G'$ increases and reaches a plateau value at $\gamma_0 \simeq 5\%$, a value comparable to $\gamma_{\rm NL}$ identified during the increasing ramp. Moreover, the plateau value of $G'$ is lower than that measured at the start of the experiment, i.e., prior to the increasing ramp: the gel that has reformed is thus weaker than the initial gel. 
The same conclusion holds for $G''$, which exhibits a more pronounced overshoot compared to the increasing ramp, yet at the same strain value $\gamma_0 \simeq \gamma_{\rm OS}$ before converging toward a plateau, whose value is lower than the value of $G''$ measured at rest, before the start of the experiment. Concomitantly $q_{\rm max}$ decreases, suggesting that large CB clusters tend to reform, leading to a recovery of the sample elastic properties. 
However, in stark contrast with the trend observed upon increasing strain amplitude, $q_{\rm max}$ follows a different path and reaches a plateau value $q_{\rm max} \simeq 18.10^{-3}~\rm nm^{-1}$ at $\gamma_0 \simeq 1\%$, a value that is significantly larger than the initial $q_{\rm max} \simeq 14.10^{-3}~\rm nm^{-1}$ [see Fig.~\ref{fig:rheo_saxs_polym}(b)]. Moreover, a similar hysteretic path is reported for the width $q_2/q_1$ of the peak in the Kratky plot and for the area $Q_{\delta q}$ upon decreasing the strain amplitude [see Fig.~\ref{fig:rheo_saxs_polym}(c,d)].
Such a hysteresis in both the scattering data and the rheological behavior strongly suggests that the original microstructure of the insulating CB-CMC hydrogel does not fully reform over the time of the experiment. Such a phenomenon may have two different physical origins: the first relates to the spatial distributions of the CB particles and clusters. While being brought closer under large-amplitude strain oscillations, some particles and/or clusters might form larger aggregates that do not break upon decreasing the strain amplitude. Second, we cannot rule out the possibility that CB clusters present at the start of the experiment may be broken up upon increasing the strain amplitude. Such broken clusters are then scattered under shear and remain as such in the polymer matrix upon decreasing $\gamma_0$, yielding different rheological properties.

\section{\label{sec:Discussion} Discussion and Conclusion}

Let us now summarize the key findings of this paper. We have used static USAXS measurements to confirm the dual nature of CB-CMC hydrogels, previously identified through electrical impedance spectroscopy and rheometry \cite{Legrand:2023}. A key novelty of the present contribution is the quantitative assessment of relative changes in CB microstructure across different hydrogel compositions. Our results reveal that the polymer content has the most impact on the CB cluster size distribution, and also suggest that, in insulating samples, CB is present both as individual particles as well as clusters heterogeneously distributed within the polymer matrix.    

We have characterized the yielding transition of CB-CMC hydrogels under LAOStrain by examining the response of more than a hundred different compositions. A central outcome of this experimental work has been to demonstrate that key features of the LAOStrain response --specifically, the yield strain $\gamma_y$ and the strain at which $G''$ shows a peak, $\gamma_{\rm OS}$-- follow an essentially binary behavior. When plotted as a function of the polymer-to-colloid ratio $r$, these observables are clearly separated by a critical ratio $r_c$, previously identified from linear viscoelastic and impedance measurements. The critical ratio $r_c$ thus emerges as a robust parameter separating both the linear and non-linear properties of conductive and insulating CB-CMC hydrogels. While electrically conductive hydrogels ($r<r_c$) display rather well-defined properties, insulating hydrogels ($r>r_c$) show much broader variability, likely due to a heterogeneous microstructure featuring individual CB particles as well as clusters. The size distribution of these clusters appears broad, dependent on sample preparation, and shear sensitive, as evidenced by the hysteresis unraveled in the structural and mechanical properties of insulating hydrogels under LAOStrain.   

Conductive and insulating CB-CMC hydrogels also differ markedly in their yielding process. On the one hand, conductive CB-CMC hydrogels, which exhibit a glassy-like, frequency-independent linear viscoelastic response \cite{Legrand:2023}, undergo a progressive yielding transition with $\gamma_{\rm NL} < \gamma_{\rm OS} <\gamma_y $. The overshoot in $G''$ corresponds to the failure of the percolated CB network, as evidenced by the concurrent decrease in the sample electrical conductivity. This observation aligns with the picture proposed by Donley \textit{et al}.~\cite{Donley:2020} in which the peak of the overshoot in $G''$ occurs at a strain amplitude where the unrecoverable deformation of the sample becomes dominant. In that framework, the yielding transition is not an abrupt phenomenon but a \textit{continuous} process, as confirmed by the progressive evolution of the electrical properties. Moreover, the CB network fails at length scales larger than those accessible to TRUSAXS, hence pointing to a spatially localized failure into large, undamaged clusters, reminiscent of solid–fluid coexistence and brittle-like failure previously reported in LAOStress and fatigue tests of CB gels in mineral oil \cite{Perge:2014,Gibaud:2016}.
However, in contrast to the latter case, further increasing the strain amplitude leads to a striking rise in conductivity of the fluidized CB-CMC hydrogel --up to twice its initial value-- due to the formation of a transient, shear-induced percolated CB network of CB clusters. This \textit{dynamic} structure rapidly relaxes upon cessation of shear and differs from the reversible gelation seen in colloid–polymer mixtures with polymer bridging \cite{Kamibayashi:2008}, since the conductivity increase here is due to direct CB-CB contacts rather than bridging interactions.

On the other hand, insulating CB-CMC hydrogels characterized by a frequency-dependent viscoelastic spectrum that obeys a time-concentration superposition principle \cite{Legrand:2023}, display a distinct yielding sequence where the peak in $G''$ is located beyond the yield point, i.e., $\gamma_{\rm NL}<\gamma_y<\gamma_{\rm OS}$. In these samples, the yield point likely reflects the disruption of a sufficient number of CB-mediated cross-links, entailing that $G'<G''$, while the subsequent overshoot in $G''$ reflects the non-linear response of the physically cross-linked polymer network, as revealed by its sensitivity to frequency (see Appendix~\ref{Appendix:ImpactFreq}). However, at this stage, it remains unclear in which proportion these two processes affect the unrecoverable strain associated with the shear-induced solid-to-liquid transition \cite{Donley:2020}. Additionally, the yielding process coincides with an overall decrease in the average CB cluster size, which most likely involves the shear-induced failure of these isolated clusters trapped in the polymer matrix. 

In conclusion, our findings shed new light on the non-linear mechanical response and microscopic dynamics of colloid-polymer hydrogels, whose solid-like microstructure arises either from a percolated colloidal network or from physically cross-linked polymers, depending on their composition. These two different microstructural architectures lead to distinct yielding behaviors governed by a single control parameter, namely the polymer-to-colloid ratio. This compositional tunability offers valuable design flexibility for applications ranging from injectable conductive gels to responsive soft actuators. In particular, the emergence of a transient conductive state under shear in CB-CMC hydrogels points to new opportunities for designing strain-sensitive materials. Such behavior may prove useful in flow battery slurries \cite{Gordon:2020,Burdette:2020,Reynolds:2022}, where dynamic percolation under flow is essential, or in shear-responsive drug delivery systems based on colloid-polymer architectures \cite{Wang:2021}. More broadly, our results advance the understanding of yielding in heterogeneous gels and motivate further investigations to reveal the link between microstructural dynamics and macroscopic rheology in a wide range of systems with composition-dependent percolation pathways.

\begin{acknowledgments}
The authors acknowledge the European Synchrotron Radiation Facility (ESRF) for provision of synchrotron radiation facilities under proposal number SC-5515 and thank T.~Narayanan for assistance and support in using beamline ID02, as well as E.~Freyssingeas for technical support during the rheo-TRUSAXS experiments. The authors also acknowledge the contribution of SFR Santé Lyon-Est (UAR3453 CNRS, US7 Inserm, UCBL) facility: CIQLE (a LyMIC member), especially S.~Karpati, for their help with the TEM observations. The authors thank J.~Blin, R.~Cardinaels, A.~Colin, P.~Coussot, A.-C. Genix, J.J.~Richards, and J.~Vermant for fruitful discussions.
This work was supported by the LABEX iMUST of the University of Lyon (ANR-10-LABX-0064), created within the ``Plan France 2030" set up by the French government and managed by the French National Research Agency (ANR). This research was supported in part by grant NSF PHY-2309135 to the Kavli Institute for Theoretical Physics (KITP). 
\end{acknowledgments}

\section*{Data Availability Statement}

The data that support the findings of this study are available from the corresponding author upon reasonable request. The data obtained at the European Synchrotron Radiation Facility (ESRF) are available at DOI: 10.15151/ESRF-ES-1465100229


\appendix

\section{Experimental limits of the rheo-electric setup} \label{appendix:rheoelec}

\subsection{Impact of the wire on rheological measurements}
\label{appendix:rheoelec1}
\begin{figure}[!ht]
    \centering
    \includegraphics[width=0.8\linewidth]{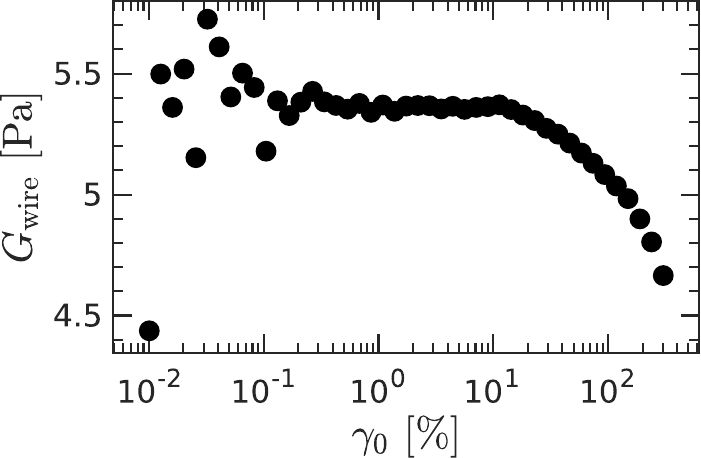}
    \caption{Effective elastic modulus $G_{\rm wire}$ measured with the rheo-electric setup in the absence of any sample, as a function of the strain amplitude $\gamma_0$ during an oscillatory shear at $\omega=2\pi~\rm rad.s^{-1}$. This elastic modulus arises from the finite bending rigidity of the wire.}
    \label{fig:rheo_elec_check_wire}
\end{figure}

The copper wire closing the electrical circuit results in a residual torque that may affect mechanical measurements. This contribution was determined by performing a strain sweep at $\omega=2\pi ~\rm rad.s^{-1}$ with an empty shear cell. The effective elastic modulus $G_{\rm wire}$ measured as a function of the strain amplitude $\gamma_0$ and reported in Fig.~\ref{fig:rheo_elec_check_wire} ranges between $4.4~\rm Pa$ and $5.8~\rm Pa$, and decreases for increasing strain amplitude as the wire unfolds. Note that the exact contribution of the wire rigidity to the rheological measurements actually depends on the manner in which the wire is attached to the rotor, which varies from one experiment to the other. This technical limitation requires that only CB-CMC hydrogels, whose viscoelastic moduli are much larger than $5~\rm Pa$, are studied.

\begin{figure*}[!th]
    \centering
    \includegraphics[width=0.8\linewidth]{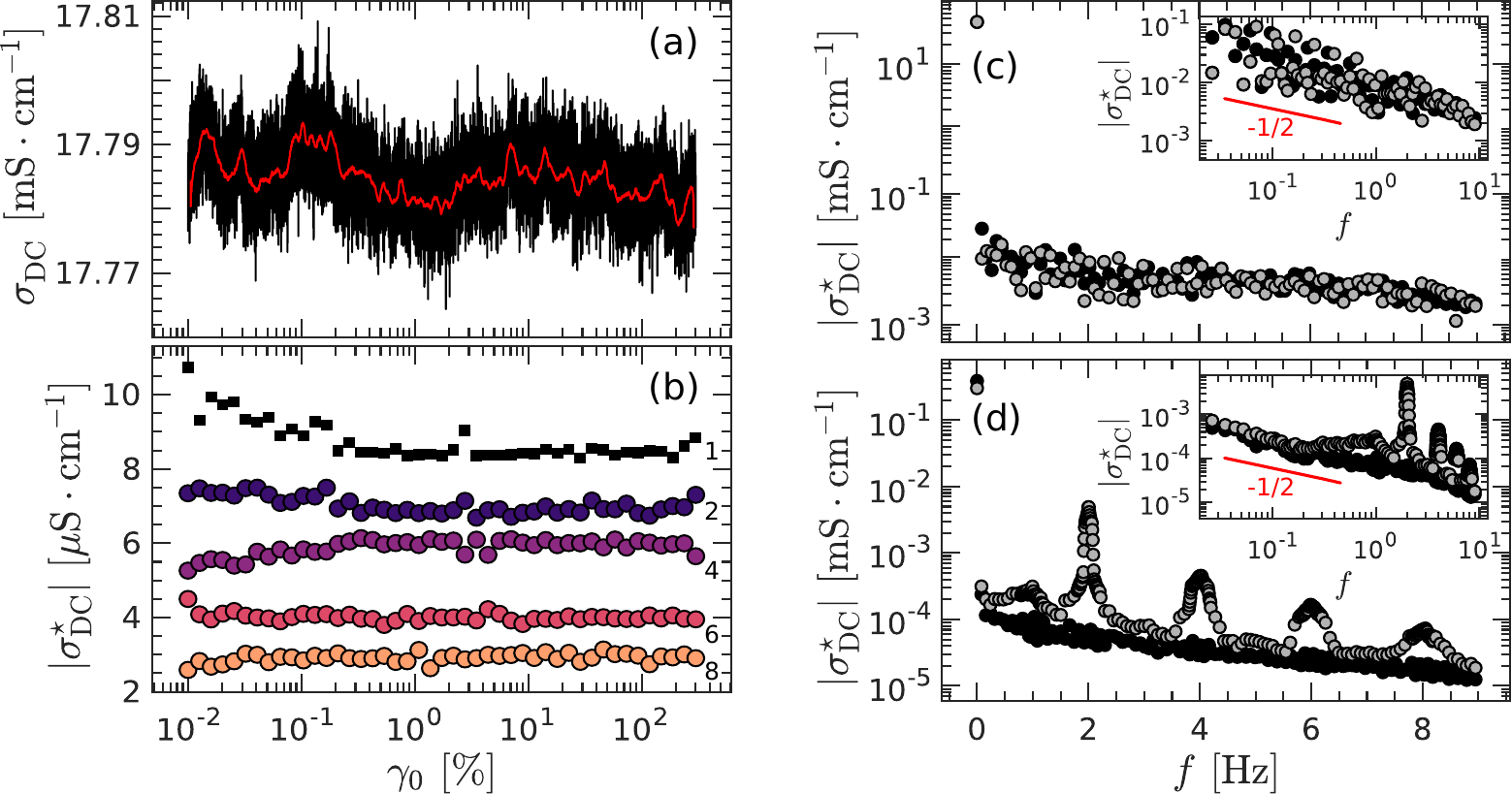}
    \caption{(a) Electrical conductivity $\sigma_{\rm DC}$ measured on the EGaIn alloy during a LAOStrain experiment. The raw data are plotted in black, while the red curve corresponds to a moving average over 1~s, i.e., 100 points. (b) Harmonic amplitudes $ | \sigma_{\rm DC, p}^\star | =|\sigma_{\rm DC}^\star(f= p \omega/2\pi)|$ vs.~$\gamma_0$ for $p=1,\, 2,\, 4,\, 6,\, \rm and\, 8$, as indicated on the graph, extracted from the spectral analysis of the raw data shown in (a), as a function of the strain amplitude $\gamma_0$. Note the change in units, i.e., the amplitude of the harmonics is 3 orders of magnitude smaller than the zero-frequency component. (c) Spectrum of $\sigma_{\rm DC}$ at small oscillation amplitude  ($\gamma_0 \simeq 0.01\%$, black) and at large oscillation amplitude $\gamma_0 \simeq 300\%$ (grey). Inset: same data in logarithmic scales. The red line shows a power-law with an exponent of $-1/2$, which corresponds to $|\sigma_{\rm DC}^\star|^2 \sim 1/f$, i.e., a power spectrum compatible with pink noise. (d) Same analysis as in (c) for a CB-CMC conductive hydrogel ($c_{\rm CMC} = 0.01\%$ and $x_{\rm CB} = 6 \%$, $r<r_c$). Analysis performed on the data reported in Fig.~\ref{fig:rheo_elec_coll} in the main text. }
    \label{fig:rheo_elec_noise}
\end{figure*}

\subsection{Impact of the wire deformation on electrical measurements}
\label{appendix:rheoelec2}

During LAOStrain experiments, the wire moves, which may impact electrical measurements. We address this issue by performing time-resolved rheo-electrical measurements on EGaIn alloy (Gallium-Indium eutectic 75.5\%-24.5\% wt., ThermoFisher), a liquid metal at room temperature, which behaves as an extremely low resistor (its response is independent of the frequency) \cite{Helal:2016}. This sample is loaded into the shear cell and a LAOStrain experiment is performed at $\omega=2\pi~\rm rad.s^{-1}$ [see Fig.~\ref{fig:rheo_elec_noise}(a,b)]. Although current fluctuations are visible, they are not correlated with the strain amplitude. This indicates that the electrical measurements can be safely assumed to be independent of the mechanical protocol. Moreover, the magnitude of the noise decays roughly as a pink noise with $|\sigma_{\rm DC}^\star|^2 \sim 1/f$, for both the reference liquid metal and the probed CB-CMC hydrogels [see Fig.~\ref{fig:rheo_elec_noise}(c,d)].  
We conclude that the wire only adds a resistive component to the electrical circuit. The typical value of this resistance is estimated to be around 2~$\Omega$ at most for all frequencies \cite{legrand:tel-04768650}.

\section{Impact of oscillation frequency on the LAOStrain response of insulating CB-CMC hydrogels} 
\label{Appendix:ImpactFreq}

\begin{figure}[!t]
    \centering
\includegraphics[width=0.9\linewidth]{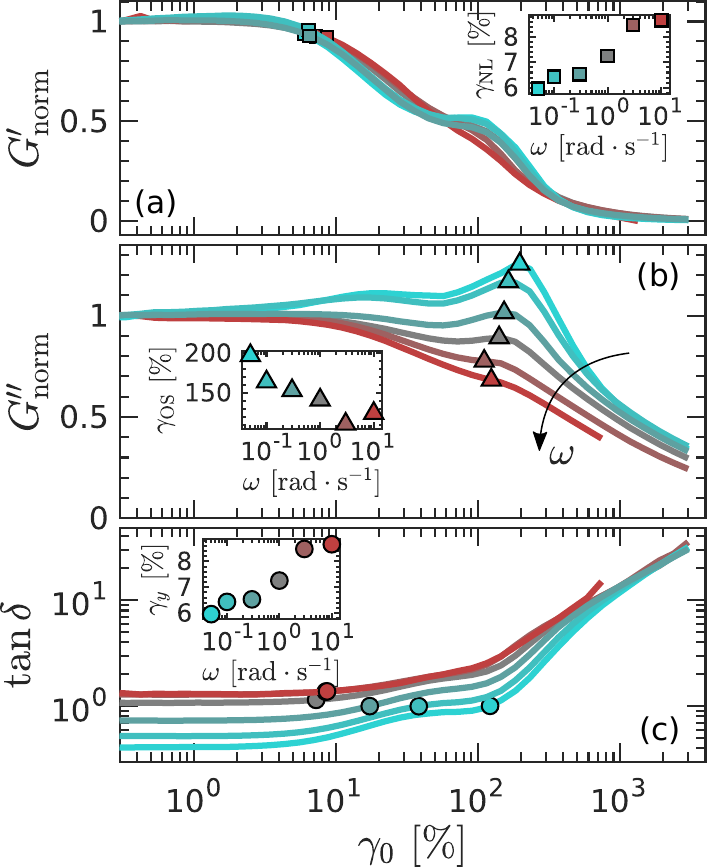}
    \caption{Impact of oscillation frequency on the LAOStrain response of insulating CB-CMC hydrogels ($c_{\rm CMC} = 2\%$ and $x_{\rm CB} = 8\%$, $r>r_c$). (a) Normalized elastic modulus $G'_{\rm norm}$, (b) normalized viscous modulus $G''_{\rm norm}$, and (c) their ratio $\tan \delta = G''_{\rm norm} / G'_{\rm norm}$ as a function of the strain amplitude $\gamma_0$ for experiments conducted at different frequencies $\omega = (0.05,\, 0.01,\, 0.3,\, 1,\, 3$ and 10) $\times 2\pi~\rm rad.s^{-1}$, going from blue to red colors as the frequency is increased. 
    Each point is averaged over two cycles, and $G'$ and $G''$ are normalized by their respective values in the linear regime.} 
\label{fig:CB_CMC_non_lin_strain_sweep_different_frequencies}
\end{figure}

The LAOStrain response of insulating CB-CMC hydrogels -- such as the one shown in Fig.~\ref{fig:strain_sweep_CB8}(b) in the main text-- exhibits a significant dependence on the frequency at which the strain sweep is performed. To illustrate this frequency dependence, Fig.~\ref{fig:CB_CMC_non_lin_strain_sweep_different_frequencies} reports the results of six strain sweep experiments performed on the same sample, at different frequencies ranging between $0.3~\rm rad.s^{-1}$ and $60~\rm rad.s^{-1}$. These frequencies are chosen to encompass the characteristic crossover frequency of the hydrogel, $\omega_0=1~\rm rad.s^{-1}$, which marks a transition in the linear viscoelastic response. Specifically, for $\omega < \omega_0$, $G' \gg G''$, and $G'$ exhibits a plateau that reflects the linear viscoelastic response of the CMC network cross-linked by CB, whereas for $\omega > \omega_0$, $G' \gtrsim G''$, and $G'$ increases following a power-law behavior, indicative of viscoelasticity dominated by the polymer and solvent contributions(see the linear viscoelastic spectrum reported in Fig.~6 in ref.~\cite{Legrand:2023}). 

For $\omega >\omega_0$, both $G'$ and $G''$ decrease monotonically with increasing strain amplitude [Fig.~\ref{fig:CB_CMC_non_lin_strain_sweep_different_frequencies}(a) and \ref{fig:CB_CMC_non_lin_strain_sweep_different_frequencies}(b)]. In contrast, at lower frequencies $\omega <\omega_0$, $G'$ develops a two-step decrease with a plateau at $\gamma_0 \simeq 100\%$, while $G''$ displays an overshoot characterized by a peak at $\gamma_0 \simeq 100\%$. Notably, the amplitude of this $G''$ overshoot increases with decreasing frequency. 
These results support the interpretation that the overshoot in $G''$ is a signature of the yielding of the polymer network crosslinked by CB particles. This interpretation is consistent with previous LAOStrain experiments performed on acid-induced CMC hydrogels that are physically crosslinked in the absence of CB, which show a qualitatively similar $G''$ overshoot behavior \cite{Legrand:2025}.

\section{Ratio of power-law exponents $\nu'/\nu''$ vs. polymer-to-particle ratio $r$}
\label{Appendix:rationu}

\begin{figure}[!h]
    \centering
    \includegraphics[width=0.9\linewidth]{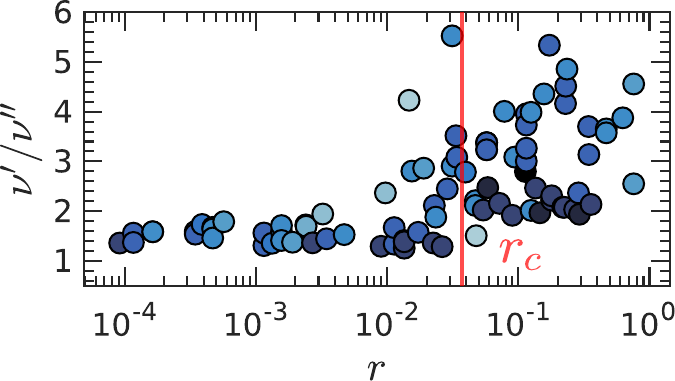}
    \caption{Ratio $\nu'/\nu''$ of the two exponents characterizing the power-law decays of $G'$ and $G''$ beyond the yield point. The ratio is computed with the data reported in Figs.~\ref{fig:strain_sweep_analyzed}(e) and \ref{fig:strain_sweep_analyzed}(f) in the main text. Colors indicate the CB content: darker colors correspond to larger $x_{\rm CB}$ [see color bar in Fig.~\ref{fig:strain_sweep_analyzed}(c)].}
    \label{fig:ratio_nu}
\end{figure}

Figure~\ref{fig:ratio_nu} displays the ratio $\nu'/\nu''$ of the two power-law exponents reported in Figs.~\ref{fig:strain_sweep_analyzed}(e) and \ref{fig:strain_sweep_analyzed}(f). Conductive CB-CMC hydrogels ($r<r_c$) share a constant value $\nu'/\nu'' \simeq 1.5$, insensitive to the sample composition. In contrast, insulating gels ($r>r_c$) exhibit larger values, with $\nu'/\nu'' \gtrsim 2$. Moreover, the ratio increases for increasing values of $r$, with significant variations with the sample composition.

\section{Relaxation of the electrical conductivity following a LAOStrain sweep}
\label{Appendix:Relaxation}

\begin{figure}[!h]
    \centering
    \includegraphics[width=0.9\linewidth]{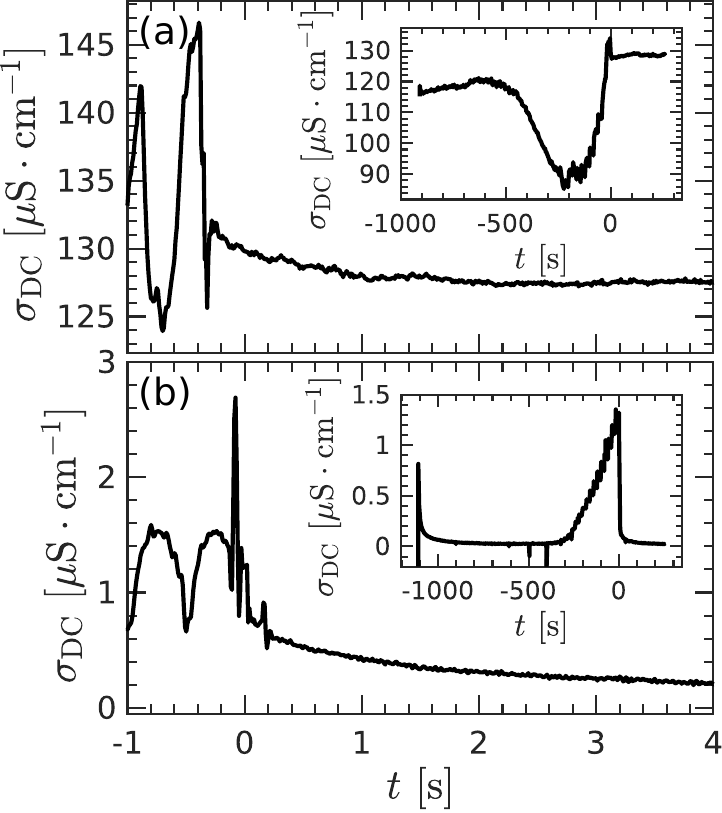}
    \caption{DC electrical conductivity $\sigma_{\rm DC}$ as a function of time $t$ for (a) a conductive CB-CMC hydrogel with $c_{\rm CMC} = 0.01\%$ and $x_{\rm CB} = 6\%$ and (b) an insulating CB-CMC hydrogel with $c_{\rm CMC} = 3\%$ and $x_{\rm CB} = 8\%$. The origin of time $t = 0$ is set at the end of the LAOStrain experiment. The insets show the full data sets, which were smoothed over $0.2~\rm s$ for clarity.}
    \label{fig:relax}
\end{figure}

The relaxation dynamics of the conductivity following flow cessation after a LAOStrain experiment conducted on a conductive and an insulating CB-CMC hydrogel is shown in Fig.~\ref{fig:relax}. In both cases, the relaxation of the DC electrical conductivity occurs typically within $1~\rm s$ following flow cessation. 

\section{Fluctuations of conductivity measured during a strain sweep on a conductive CB-CMC hydrogel}
\label{Appendix:ElecVariation}

\begin{figure}[!h]
    \centering
    \includegraphics[width=0.95\linewidth]{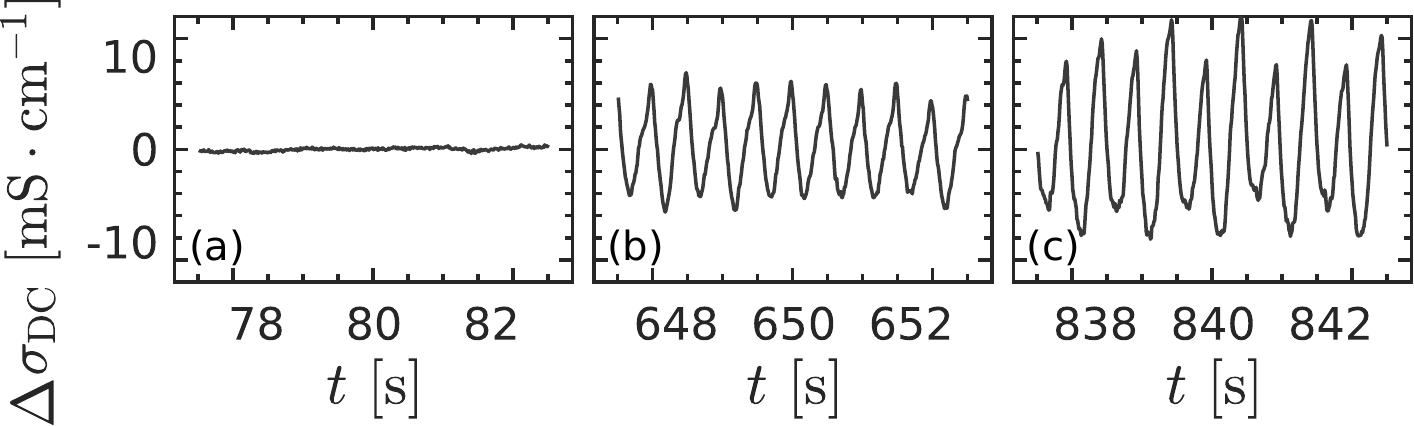}
    \caption{Fluctuations of conductivity $\Delta \sigma_{\rm DC}(t)=\sigma_{\rm DC}(t)-\langle\sigma_{\rm DC}\rangle$ observed in the raw data of the DC electrical conductivity as a function of time for different strain values along the strain sweep: (a) $\gamma_0 = 0.03 \%$, (b) $\gamma_0 = 20 \%$, and (c) $\gamma_0 = 200 \%$.$\langle\sigma_{\rm DC}\rangle$ denotes the average of $\sigma_{\rm DC}(t)$ over each time window. The three strain values are marked by red stars in Fig.~\ref{fig:rheo_elec_coll}(c) in the main text. }
    \label{fig:zoom_conducti}
\end{figure}

Figure~\ref{fig:zoom_conducti} illustrates the level of fluctuations in the DC electrical conductivity recorded during the LAOStrain experiments reported in Fig.~\ref{fig:rheo_elec_coll}(c) in the main text. The conductivity is reported as a function of time for the three strain values indicated by a red star in Fig.~\ref{fig:rheo_elec_coll}(c). The fluctuations are highlighted by removing the mean value of $\sigma_{\rm DC}$ over each time window, which displays exactly 10 periods.

\section{Impact of the mechanical oscillation on the electrical conductivity}
\label{Appendix:calc}

\begin{figure}[!th]
    \centering
    \includegraphics[width=0.95\linewidth]{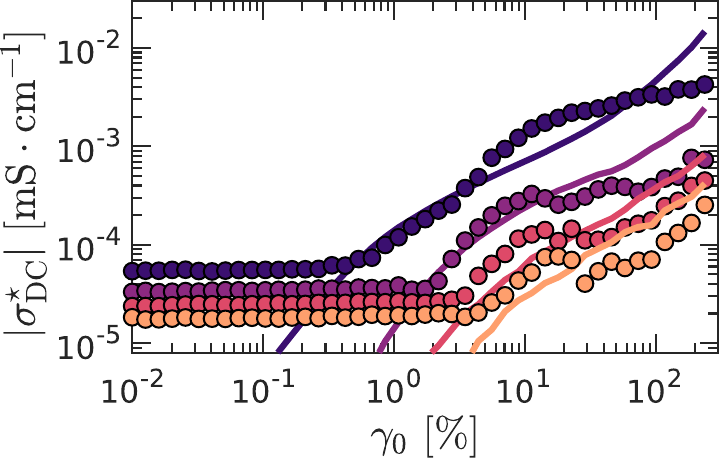}
    \caption{Harmonic amplitudes $|\sigma_{\rm DC, p}^\star|=|\sigma_{\rm DC}^\star(f= p \omega/2\pi)|$ vs.~$\gamma_0$ with $p= 2$, 4, 6, and 8 as indicated on the graph. Same data as those reported in Fig.~\ref{fig:rheo_elec_coll}(e). The curves correspond to the best fits of the data with Eq.~\eqref{eq:energy}, with $c^\prime_1 = 40 \ \rm mS.cm^{-1}.Pa^{-1}$ ($\gamma_0$ is expressed in strain units)}
    \label{fig:Energy}
\end{figure}

\begin{figure*}[!th]
    \centering
    \includegraphics[width=0.95\linewidth]{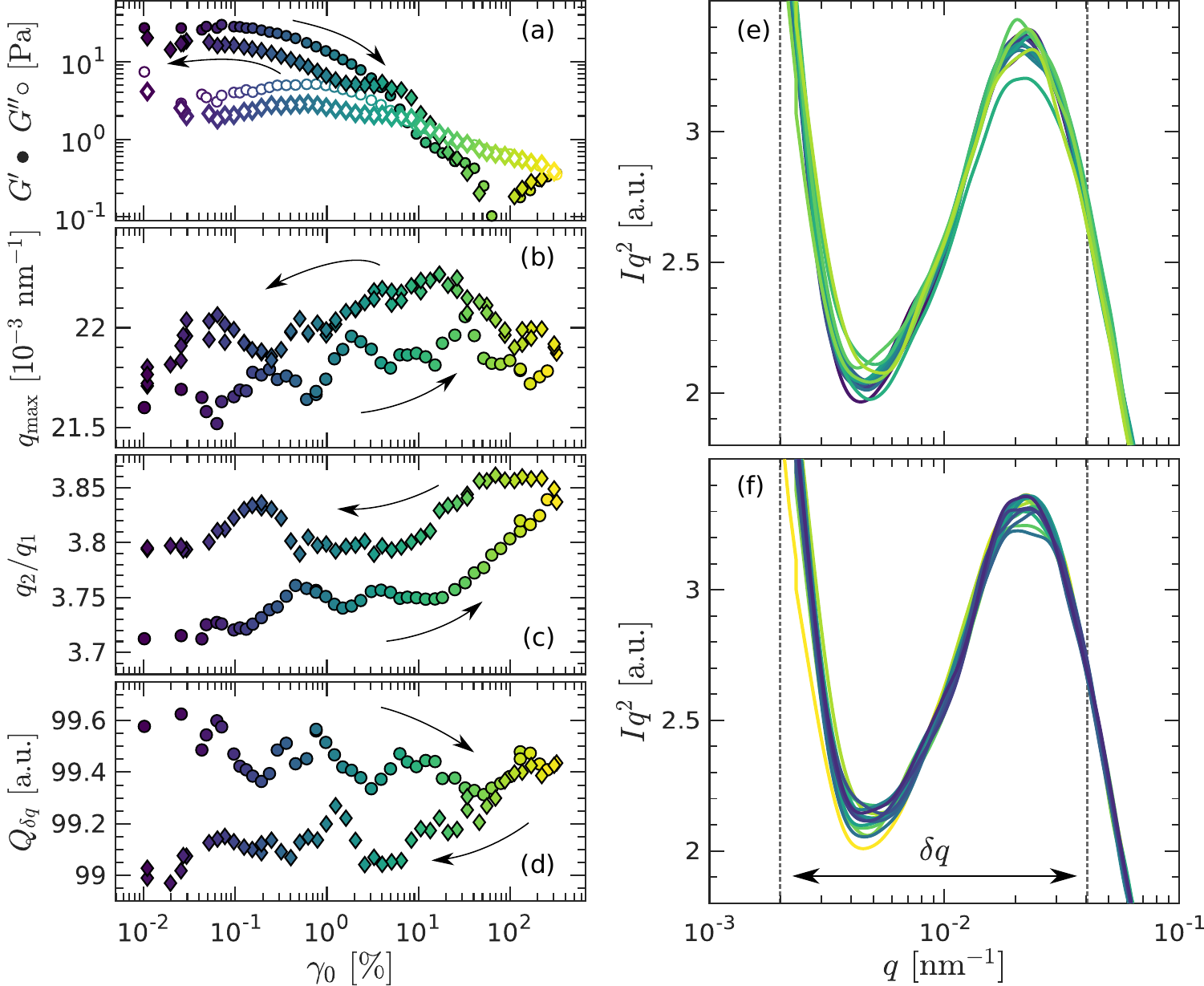}
    \caption{LAOStrain experiment performed using rheo-TRUSAXS on a conductive CB-CMC hydrogel ($c_{\rm CMC} = 0.01\%$ and $x_{\rm CB} = 6\%$, $r<r_c$). The frequency is set to $\omega = 2\pi ~\rm rad.s^{-1}$, and the strain amplitude $\gamma_0$ is progressively increased from $0.01~\%$ to $300~\%$ (circles) and then decreased over the same range of values (diamonds), as indicated by the black arrows.
    (a) Elastic modulus $G'$ (filled symbols) and viscous modulus $G''$ (empty symbols) moduli as a function of $\gamma_0$. (b) Scattering wave vector $q_{\rm max}$ corresponding to the local maximum in the Kratky plot $Iq^2$ vs.~$q$ and plotted against $\gamma_0$. (c) Width of the peak $q_2 / q_1$ as a function of $\gamma_0$, where $q_2 = 4.10^{-2} \rm ~mm^{-1}$ and $q_1<q_2$ is the largest scattering wave vector below $q_{\rm max}$ that corresponds to $I(q_2)q_2^2$ in the Kratky plot, and indicated as colored dots. (d) Area $Q_{\delta q}$ enclosed by the Kratky plot between the two vertical dashed lines shown in (e) and (f) and plotted against $\gamma_0$. (e,f) Kratky plots $Iq^2$ vs.~$q$ during (e) the increasing strain ramp and (f) decreasing strain ramp. Colors indicate the strain amplitude $\gamma_0$, with lighter colors representing larger values of $\gamma_0$, as shown by the symbol colors in (a--d).}
    \label{fig:rheo_saxs_coll}
\end{figure*}

Here, we report on the intermediate steps for the calculation of the mechanical power $\mathcal{P} = \dot \gamma \tau $ injected into the system. Starting from the expression of the stress output Eq.~\eqref{eq:stress}, the power takes the following expression: 
\begin{align}
    \sigma_{\rm DC} = c_{0} + c_{1} \mathcal{P} + \mathcal{O}(\mathcal{P}^{2})\,.
\end{align}
The constant term $c_{0}$ should be interpreted as $\sigma_{\rm DC}$ in Fig.~\ref{fig:rheo_elec_coll}(c), i.e., as the only non-oscillating value. We shall now study the linear term in $ \mathcal{P}$: 
\begin{align}
    \mathcal{P}(t)  &= \omega \gamma_0 \tau_0 \sum_k a_k  \cos \left(  \omega t \right) \sin \left(  k \omega t + \delta_k \right) \\
    &= \frac{1}{2} \omega \gamma_0 \tau_0 \sum_k a_k \left[ \sin \left(  (k+1) \omega t + \delta_k \right) + \sin \left(  (k-1) \omega t + \delta_k \right) \right] \\
    &=  \frac{1}{2} \omega \gamma_0 \tau_0 \sum_n a_{n-1} \sin \left(  n \omega t + \delta_{n-1} \right) + a_{n+1} \sin \left(  n \omega t + \delta_{n+1} \right) \\
    &=  \frac{1}{2} \omega \gamma_0 \tau_0 \sum_n b_{n} \sin \left(  n \omega t - \phi_{n} \right) \,,
\end{align}
where we set $a_{-1} = 0$ and the amplitude $b_n$ and the phase $\phi_n$ are defined as follows:
\begin{align}
    b_{n} &= \sqrt{ a_{n+1}^2 +  a_{n-1}^2  + 2  a_{n+1} a_{n-1} \cos \left( \delta_{n+1}-\delta_{n-1} \right)} \label{eq.bn}\\
    \cos \phi_{n} &= \dfrac{a_{n+1} \cos \delta_{n+1}  + a_{n-1} \cos \delta_{n-1} }{ b_{n} }   \\  
    \sin \phi_{n} &= \dfrac{a_{n+1} \sin \delta_{n+1}  + a_{n-1} \sin \delta_{n-1} }{ b_{n} }   \,. 
\end{align}
Therefore, the term oscillating at $n\omega$ in the electrical measurement arises from the terms oscillating at $(n \pm 1) \omega$ in the stress response $\tau(t)$. The amplitudes of the harmonics of the electrical signal are linked to the non-linear rheological response by the following relation:
\begin{align}
    | \sigma_{\rm DC, n}^\star | = |\sigma_{\rm DC}^\star(f= n \omega/2\pi)|  &= \frac{1}{2} c_1 \omega \gamma_0 \tau_0   b_n  \\
     &= \frac{1}{2} c_1 |G^\star| \omega \gamma_0^2 b_n\,,
 \label{eq:power}
\end{align}
where $c_1$ is a constant that neither depends on $\gamma_0$ nor $n$. 

Note that, rather than focusing on the instantaneous power $\mathcal{P}$, one may follow the same line of thought based on the instantaneous energy input $\mathcal{E}= \gamma \tau$ by assuming that, up to second order terms,
\begin{align}
    \sigma_{\rm DC} = c^{\prime}_{0} + c^{\prime}_{1} \mathcal{E} + \mathcal{O}(\mathcal{E}^{2}).
\end{align}
In that case, one gets:
\begin{align}
    \mathcal{E}  &= \gamma_0 \tau_0 \sum_k a_k  \sin \left(  \omega t \right) \sin \left(  k \omega t + \delta_k \right) \\
    &=  \frac{1}{2} \gamma_0 \tau_0 \sum_n b{^\prime}_{n} \cos \left(  n \omega t - \phi^\prime_{n} \right) \,,
\end{align}
where, with $a_{-1} = 0$, the amplitude $b^\prime_n$ and the phase $\phi^\prime_n$ read:
\begin{align}
    b^\prime_{n} &= \sqrt{ a_{n+1}^2 +  a_{n-1}^2  - 2  a_{n+1} a_{n-1} \cos \left( \delta_{n+1}-\delta_{n-1} \right)} \\
    \cos \phi^\prime_{n} &= \dfrac{a_{n+1} \cos \delta_{n+1}  + a_{n-1} \cos \delta_{n-1} }{ b^\prime_{n} }   \\  
    \sin \phi^\prime_{n} &= \dfrac{a_{n+1} \sin \delta_{n+1}  + a_{n-1} \sin \delta_{n-1} }{ b^\prime_{n} } \,.
\end{align}
This leads to the following relation between the amplitudes:
\begin{align}
    | \sigma_{\rm DC, n}^\star | = |\sigma_{\rm DC}^\star(f= n \omega/2\pi)| &= \frac{1}{2} c^\prime_1 \gamma_0 \tau_0   b^\prime_n  \\
     &= \frac{1}{2} c^\prime_1 |G^\star| \gamma_0^2   b^\prime_n\,. \label{eq:energy}
\end{align}
Note that Equations~\eqref{eq:power} and \eqref{eq:energy} both lead to $| \sigma_{\rm DC, n}^\star | \propto |G^\star| \gamma_0^2  {a_{n-1}} $ under the assumption $|a_{n+1}| \ll | a_{n-1}|$. Therefore, both approaches are equivalent in the weakly non-linear regime.

As shown in Fig.~\ref{fig:Energy}, the harmonic amplitudes reported in Fig.~\ref{fig:rheo_elec_coll}(e) can be fitted using Eq.~(\ref{eq:energy}) with $c^\prime_1 \simeq =40 \pm 5\ \rm mS.cm^{-1}.Pa^{-1}$ ($\gamma_0$ is expressed in strain units). We observe that the agreement between the model and the data is satisfying for harmonics $2$ and $4$. However, the agreement for harmonics $6$ and $8$ is not as good as the one obtained with the model based on the instantaneous power $\mathcal{P}$, thanks to Eq.~\eqref{eq:power}, as reported in Fig.~\ref{fig:rheo_elec_coll}(e). 

\section{Insights from rheo-TRUSAXS on a conductive CB-CMC hydrogel}
\label{Appendix:TRUSAXSconductive}

Rheo-TRUSAXS experiments were performed on a conductive CB-CMC hydrogel. The result of a strain sweep of increasing amplitude immediately followed by a ramp of decreasing amplitude over the same range of strains is reported in Fig.~\ref{fig:rheo_saxs_coll}.

The response of the hydrogel to increasing strain amplitudes, shown in Fig.~\ref{fig:rheo_saxs_coll}(a), is consistent with that obtained in other geometries and reported in Figs.~\ref{fig:strain_sweep_CB8}(a) and \ref{fig:rheo_elec_coll}(a). The scattering intensity remains roughly the same, and no significant changes are observed either in the position of the peak in $Iq^2$, nor in its width. The same conclusion holds upon decreasing the strain amplitude, which suggests that shear-induced microstructural changes occur at length scales inaccessible to the USAXS window, i.e., above a few microns at least. Additionally, if shear would impact the average distance between CB particles, it would appear in the scattering data. Therefore, the lack of change in the scattering data also suggests that the hydrogel breaks into large-scale pieces that subsequently remain unaffected by shear, as discussed in the main text.


%

\end{document}